\documentclass[12pt]{revtex4}
\usepackage{epsfig}

\begin{document}

\title{Stochastic Variational Method as Quantization Scheme I: Field
Quantization of Complex Klein-Gordan Equation}
\author{T. Koide}
\email{tomoikoide@gmail.com,koide@if.ufrj.br}
\author{T. Kodama}
\email{kodama.takeshi@gmail.com,tkodama@if.ufrj.br}
\affiliation{Instituto de F\'{\i}sica, Universidade Federal do Rio de Janeiro, C.P.
68528, 21941-972, Rio de Janeiro, Brazil}

\begin{abstract}
Stochastic Variational Method (SVM) is the generalization of the variation
method to the case with stochastic variables. In the series of papers, we
investigate the applicability of SVM as an alternative field quantization
scheme. Here, we discuss the complex Klein-Gordon equation. In this scheme,
the Euler-Lagrangian equation for the stochastic fields leads to the
functional Schr\"{o}dinger equation, which in turn can be interpreted as the
ideal fluid equation in the functional space. We show that the Fock state
vector is given by the stationary solution of these differential equations
and various results in the usual canonical quantization can be reproduced,
including the effect of anti-particles. The present formulation is a
quantization scheme based on commutable variables, so that there appears no
ambiguity associated with the ordering of operators, for example, in the
definition of Noether charges. 
\end{abstract}

\pacs{03.70.+k,11.10.Ef,05.40.-a}
\maketitle

\section{introduction}

Variational approach plays conceptually a fundamental role in elucidating
the structure of Classical Mechanics, unifying the origin of dynamics and
the relation between symmetries and conservation laws. This is the reason
that it is referred to as ``principle" rather than the simple re-expression
of Newton's equation of motion. In quantum mechanics, such an aspect is
somewhat lost. As a matter of fact, quantum dynamics is given by the Schr%
\"{o}dinger equation, and symmetries and conservation laws are formulated as
transformation groups in the Hilbert space. Of course, although it is
possible to write down an action which reproduces the Schr\"{o}dinger
equation, such a re-expression does not provide further insight into the
structure of quantum mechanics. Rather, the variational approach in quantum
mechanics is usually used for practical purposes. Inversely, as seen in the
path integral formulation, quantum mechanics is considered to give an
explanation for the origin of classical variational principle in the sense
that the classical optimized path gives the dominant contribution in the
transition amplitude.

It was, however, shown by K. Yasue \cite{yasue} around three decades ago
that the fundamental concepts of quantum mechanics can be formulated in
terms of the variational principle, generalizing classical variables to the
corresponding stochastic one, whose stochastic property is known as
Bernstein (reciprocal) process \cite{bern}. This new approach is called the
stochastic variational method (SVM) and was originally introduced to
reformulate Nelson's stochastic quantization \cite{nelson} in terms of the
variational principle. In this formulation, classical mechanics and quantum
mechanics are unified under the principle of the optimization of actions in
such a way that the Schr\"{o}dinger equation is derived by employing SVM to
the action which reproduces Newton's equation of motion.

The applicability of SVM is not restricted to the derivation of quantum
mechanics. It has been shown that the Navier-Stokes equation \cite%
{kk1,kk2,koide}, the Gross-Pitaevskii equation \cite{morato,kk2} and the
Kostin equation \cite{misawa} can be formulated in the framework of SVM. The
Noether theorem \cite{misawa3}, the uncertainty relation \cite{kk3} and the
applications to the many-body particle systems \cite{morato} and continuum
media \cite{kk1,kk2} are also possible to be formulated in this scheme. The
foundation of this kind of stochastic calculus has been investigated by
various authors \cite%
{dav,guerra,zamb,hase,marra,jae,pav,rosen,wang,naga,cre,arn,kappen,eyink,gomes,hiroshima,serva,yamanaka}%
. Considering these interesting aspects of SVM, it is worth exploring the
applicability to more complex systems.

The purpose of this series of papers is to investigate the applicability of
SVM to the quantization of bosonic fields. This paper is Part I of this
series and we investigate the field quantization of the complex Klein-Gordon
equation. By using the result established in this paper, we will discuss the
possibility of the gauge field quantization without the gauge fixing
condition in Part II. 

At first glance, the applicability to the complex Klein-Gordon equation
seems to be trivial because it is known that a free field can be regarded as
the ensemble of harmonic oscillators as usually done in the canonical
quantization. However, there are several points to which we have to pay
attention when the SVM formulation is generalized to field systems. For
example, 1) the field quantization based on SVM should be formulated on both
of the field $\phi \left( \mathbf{x},t\right) $ and its Fourier transform $%
\phi (k,t)$ on an equal footing, 2) we need to introduce a singular spatial
corrections for stochastic variables so as to exist a well-defined continuum
limit, 3) the SVM formulation should treat anti-particle appropriately. 

In this paper, we show that it is possible to formulate the SVM field
quantization satisfying these points. We further find that the difference of
the SVM quantization from the canonical quantization appears in the
definition of the Noether charge. If we take naively the hermitian form for
the Noether charge operator of the complex Klein-Gordon equation, it
contains a constant divergent term. Since any constant term does not affect
dynamics, this term is simply ignored or removed by the normal ordering
product. There is no such an ambiguity in SVM because stochastic
variables are commutable. As is shown in the present paper, the Noether
charge of the complex Klein-Gordon equation can be defined straightforwardly
without introducing the normal ordering product. This is an advantage of our
quantization scheme. 

This paper is organized as follows. Before discussing the complex
Klein-Gordon equation, we first examine, in Sec. II, the quantization of the
classical string to illustrate various notations and definitions needed in
the SVM field quantization. Then the quantization of the complex
Klein-Gordon equation is discussed in Sec. III. In Sec. IV, we discuss the
determination of the intensity of the noise. Other aspects of SVM such as
anti-particles, propagator, the application to the case of the imaginary
mass and the continuum limit are discussed in Sec. V. Section VI is devoted
to the concluding remarks.

\section{Classical string}

As was mentioned in the introduction, we need to show that the SVM
quantization can be formulated not only with the field variable but also
with its Fourier transform. To discuss this, we need various notations and
definitions. To make these definitions clear, it is better to consider the
quantization of a simple system, the one-dimensional string. We show how the
two equivalent representations, the $x-$ and $k-$representations are
introduced. 

\subsection{SVM quantization: $x$-space}

The classical dynamics of the string with a small amplitude $y=y(x,t)$ is
described by the following equation of motion, 
\begin{equation}
\frac{\partial ^{2}y}{\partial t^{2}}-\frac{T}{\sigma } \frac{\partial ^{2}y%
}{\partial x^{2}}=0,  \label{st-1}
\end{equation}
where $\sigma $ and $T$ are the line density of mass and tension,
respectively. For the sake of simplicity, we consider the spatially
one-dimensional system. The corresponding action is then given by 
\begin{equation}
I [y]=\int^{t_f}_{t_i} dt\int^L_0 dx\ \left[ \frac{\sigma }{2}\left( \frac{%
\partial y}{\partial t}\right) ^{2}-\frac{T}{2}\left( \frac{\partial y}{%
\partial x}\right) ^{2}\right] ,  \label{classical_string}
\end{equation}
where $t_i$ and $t_f$ denote the initial and final times, respectively and $%
L $ is the total length of the string.

In SVM, the dynamical variable $y(x,t)$ is extended to the stochastic
variable $\widehat{y}(x,t)$. In the following argument, we use the accent
symbol $\widehat{~~}$ to denote stochastic variables. The time evolution and
distribution of the stochastic variables are not smooth and hence the time
and space derivatives should be modified from the usual classical ones.

To define the quantity corresponding to the space derivative, we discretize 
the space into a lattice. Let us divide $L$ by $N$ grid points, where $N$ is
the odd integer. The grid interval is given by $\Delta x=L/N$. We observe
the string dynamics on these grid points. Then the amplitude $y$ on the grid
points are expressed as a vector 
\begin{equation}
y(x,t)=\left( 
\begin{array}{c}
y\left( x_{0},t\right) \\ 
\vdots \\ 
y\left( x_{n},t\right) \\ 
\vdots \\ 
y\left( x_{N-1},t\right)%
\end{array}%
\right) \longrightarrow \underline{y}\left( t\right) \equiv \left( 
\begin{array}{c}
y_{0}\left( t\right) \\ 
\vdots \\ 
y_{n}\left( t\right) \\ 
\vdots \\ 
y_{N-1}\left( t\right)%
\end{array}%
\right) ,
\end{equation}%
where $x_{n}=x_{0}+n\Delta x$. In the following, we use $\underline{~~}$ to
indicate a vector quantity for the amplitude configuration space. We further
adapt the periodic boundary condition as 
\begin{equation}
y_{n+N}\left( t\right) =y_{n}\left( t\right) .  \label{periodic}
\end{equation}

On this lattice, we can define two space-derivative matrices: $\nabla _{+}$
and $\nabla_{-}$, which are defined by the $N\times N$ matrices as 
\begin{equation}
\nabla _{+}=\frac{1}{\Delta x}\left( 
\begin{array}{cccccc}
-1 & 1 & 0 & \cdots & 0 & 0 \\ 
0 & -1 & 1 & 0 & \cdots & 0 \\ 
\vdots &  & \ddots &  &  & \vdots \\ 
\vdots &  &  & \ddots &  & \vdots \\ 
0 & 0 & \cdots & 0 & -1 & 1 \\ 
1 & 0 & \cdots & 0 & 0 & -1%
\end{array}%
\right) ,\ \ \nabla _{-}=\frac{1}{\Delta x}\left( 
\begin{array}{cccccc}
1 & 0 & 0 & \cdots & 0 & -1 \\ 
-1 & 1 & 0 & \cdots & 0 & 0 \\ 
\vdots &  & \ddots &  &  & \vdots \\ 
\vdots &  &  & \ddots &  & \vdots \\ 
0 & \cdots & 0 & -1 & 1 & 0 \\ 
0 & 0 & \cdots & 0 & -1 & 1%
\end{array}%
\right) ,
\end{equation}%
respectively. Note that $\nabla_{+}=-\nabla _{-}^{T}$. Using these
definitions, the matrix $\Delta _{x}$, which corresponds to the Laplacian
operator, is expressed as 
\begin{eqnarray}
\Delta _{x} &\equiv &\nabla _{-}\nabla _{+}=\nabla _{+}\nabla _{-}  \nonumber
\\
&=&\frac{1}{\left( \Delta x\right) ^{2}}\left( 
\begin{array}{cccccc}
-2 & 1 & 0 & \cdots & 0 & 1 \\ 
1 & -2 & 1 & 0 & \cdots & 0 \\ 
\vdots &  & \ddots &  &  & \vdots \\ 
\vdots &  &  & \ddots &  & \vdots \\ 
0 & \cdots & 0 & 1 & -2 & 1 \\ 
1 & 0 & \cdots & 0 & 1 & -2%
\end{array}%
\right) .  \label{matrix-laplacian}
\end{eqnarray}%
Note that these representations are constructed so as to satisfy the partial
integration formula as 
\begin{eqnarray}
\left( \underline{f}\ast \nabla _{\pm }\underline{h}\right) _{x} &=&-\left(
\nabla _{\mp }\underline{f}\ast \underline{h}\right) _{x}, \\
\left( \underline{f}\ast \Delta _{x}\underline{h}\right) _{x} &=&-\left(
\nabla _{+}\underline{f}\ast \nabla _{+}\underline{h}\right) _{x}=-\left(
\nabla _{-}\underline{f}\ast \nabla _{-}\underline{h}\right) _{x}.
\end{eqnarray}
Here we introduced the notation for the scalar product of two arbitrary $x$%
-space vectors, $\underline{f}$ and $\underline{h}$, as 
\begin{equation}
\left( \underline{f}\ast \underline{h}\right) _{x}\equiv \Delta
x\sum_{n=0}^{N-1}f_{n}h_{n}.  \label{ScalarProduct}
\end{equation}%
The factor $\Delta x$ above is introduced to reproduce the inner product of
two functions $f\left( x\right) $ and $h\left( x\right) $ as%
\begin{equation}
\left( f\ast h\right) _{x}=\int dx\ f\left( x\right) h\left( x\right) ,
\end{equation}%
in the continuum limit $\left( L,N\rightarrow \infty ,\ \Delta
x=L/N\rightarrow 0\right)$. In this limit, any $N$-vector $\underline{y}$
becomes a functional of $y(x)$.

The usual space derivative appearing in the classical action should be
expressed using these matrices. Then we assume that the space derivative is
expressed by the average of these two space derivatives $\nabla_+$ and $%
\nabla_-$ in the lattice representation. Therefore, the lattice
representation of the classical action corresponding to Eq.(\ref%
{classical_string}) is given by 
\begin{eqnarray}
I[y] &=&\int_{t_{i}}^{t_{f}}dt \left[ \frac{\sigma }{2}\left( \frac{\partial 
\underline{y}}{\partial t} * \frac{\partial \underline{y}}{\partial t}
\right)_{\Delta x} -\frac{T}{4} \left( \nabla_{+}\underline{y} * \nabla_{+}%
\underline{y}\right)_{\Delta x} -\frac{T}{4}\left( \nabla_{-}\underline{y} *
\nabla_{-}\underline{y}\right)_{\Delta x}\right]  \nonumber \\
&=&\int_{t_{i}}^{t_{f}}dt\ \left[ \frac{\sigma }{2}\left( \frac{\partial 
\underline{y}}{\partial t} * \frac{\partial \underline{y}}{\partial t}
\right)_{\Delta x} +\frac{T}{2} (\underline{y}* \Delta_{x}\underline{y}%
)_{\Delta x}\right] .  \label{LatticeAction}
\end{eqnarray}

As was mentioned, even the time derivative appearing above should be changed
when $y$ is replaced by the stochastic variable. This is discussed when the
stochastic variation is introduced.

In SVM, the dynamical variable, $\underline{y}\left( t\right) $ is extended
to a stochastic variable, 
\begin{equation}
\underline{y}\left( t\right) \rightarrow \underline{\widehat{y}}\left(
t\right) .
\end{equation}%
Moreover, as is usual with the variational method, we consider that the
stochastic process where the initial and final conditions are fixed, such as
the Bernstein (reciprocal) process \cite{bern}. To characterize this
process, we need to introduce two stochastic differential equations (SDEs)
for the amplitudes $\left\{ \widehat{y}_{n}(t),\ n=0,\cdots ,N-1\right\} $.
One is the \textit{forward} SDE which describes the forward time evolution $%
dt>0$, 
\begin{equation}
d\widehat{y}_{n}(t)=u_{n}(\underline{\widehat{y}}(t),t)dt+\sqrt{\frac{2\nu }{%
\Delta x}}dW_{n}(t),  \label{SDE_y}
\end{equation}%
where $\underline{u}=\left\{ u_{n},\ n=0,\cdots ,N-1\right\} $ represents
the velocity field to be determined as a smooth function of $\underline{%
\widehat{y}}(t)$ and $t$ as a result of the variation. The last term on the
right hand side is the noise given by the Wiener process, satisfying the
following correlation properties, 
\begin{eqnarray}
E\left[ d\underline{W}(t)\right] &=&0,  \label{correc1} \\
\hspace{-1cm}E\left[ d\underline{W}(t)d\underline{W}^{T}(t)\right] &=&|dt|\
I,  \label{correc2}
\end{eqnarray}%
where $I$ denotes the unit matrix and $E[~~]$ represents the event average
for the noise.

The other is the \textit{backward} SDE which describes the backward time
evolution $dt<0$, 
\begin{equation}
d\widehat{y}_{n}(t)=\tilde{u}_{n}(\underline{\widehat{y}}(t),t)dt+\sqrt{%
\frac{2\nu }{\Delta x}}d\tilde{W}_{n}(t).
\end{equation}%
The correlation properties of $\underline{\tilde{W}}(t)$ are the same as $%
\underline{W}(t)$, but there is no correlation between $\underline{\tilde{W}}%
(t)$ and $\underline{W}(t)$. For the sake of simplicity, we omit $\widehat{~~%
}$ for $\underline{W}(t)$ and $\underline{\tilde{W}}(t)$ .

From these SDEs, the Fokker-Planck equations are derived. We define the
amplitude configuration distribution as 
\begin{equation}
\rho (\underline{y},t)=E[\delta ^{\left( N\right) }(\underline{y}-\underline{%
\widehat{y}}(t))].  \label{density}
\end{equation}%
Then the Fokker-Planck equations obtained from the forward SDE and the
backward SDE are, respectively, given by 
\begin{eqnarray}
\partial _{t}\rho (\underline{y},t) &=&\underline{\nabla }_{y}\cdot \left( -%
\underline{u}+\frac{\nu }{\Delta x}\underline{\nabla }_{y}\right) \rho (%
\underline{y},t),  \label{fp-stri} \\
\partial _{t}\rho (\underline{y},t) &=&\underline{\nabla }_{y}\cdot \left( -%
\underline{\tilde{u}}-\frac{\nu }{\Delta x}\underline{\nabla }_{y}\right)
\rho (\underline{y},t).
\end{eqnarray}%
Here $\underline{\nabla }_{y}=(\partial /\partial y_{0},\partial /\partial
y_{1},\cdots ,\partial /\partial y_{N-1})$ and the symbol $\cdot $
represents the product of $N$ vectors as $\displaystyle \underline{A}\cdot 
\underline{B}=\sum_{n=0}^{N-1}A_{n}B_{n}$. For the two equations to be
equivalent, the following consistency condition should be satisfied, 
\begin{equation}
\underline{u}(\underline{y},t)=\underline{\tilde{u}}(\underline{y},t)+\frac{%
2\nu }{\Delta x}\underline{\nabla }_{y}\ln \rho (\underline{y},t).
\label{consistency}
\end{equation}%
See also Ref. \cite{kk2} for the corresponding argument in particle systems.

Through this condition, the unknown function $\underline{\tilde{u}}(%
\underline{y},t)$ is obtained with this condition when $\underline{u}(%
\underline{y},t)$ is given. The purpose of SVM is to determine the equation
for the remaining unknown function $\underline{u}(\underline{y},t)$ using
the variational procedure.

To implement this operation, we need to express the classical lattice action
(\ref{LatticeAction}) in terms of the stochastic variables. The problem in
this procedure is the treatment of the time derivative term. Correspondingly
to the two SDEs, at least, the two different definitions of the time
derivative are available: one is the mean forward derivative $D$ and the
other the mean backward derivative $\tilde{D}$, which are defined by 
\begin{eqnarray}
D\underline{\widehat{y}}(t) &=&\lim_{dt\rightarrow 0+}E\left[ \left. \frac{%
\underline{\widehat{y}}(t+dt)-\underline{\widehat{y}}(t)}{dt}\right\vert 
\mathcal{P}_{t}\right] , \\
\tilde{D}\underline{\widehat{y}}(t) &=&\lim_{dt\rightarrow 0+}E\left[ \left. 
\frac{\underline{\widehat{y}}(t)-\underline{\widehat{y}}(t-dt)}{dt}%
\right\vert \mathcal{F}_{t}\right] ,
\end{eqnarray}%
respectively. These are originally introduced by Nelson \cite{nelson}. Here $%
E[g(t^{\prime })|\mathcal{P}_{t}]$ denotes the conditional average of a time
sequence of stochastic variables $\left\{ g\left( t^{\prime }\right)
,t_{a}<t^{\prime }<t_{b}\right\} $ only for $t^{\prime }>t$, fixing the
values of $g\left( t^{\prime }\right) $ for $t^{\prime }\leq t$. Similarly, $%
E[g(t^{\prime })|\mathcal{F}_{t}]$ is the conditional average for the past
sequence. For the $\sigma $-algebra of all measurable events, $\{\mathcal{P}%
_{t}\}$ and $\{\mathcal{F}_{t}\}$ represent an increasing and a decreasing
family of sub-$\sigma $-algebras, respectively \cite{cre}. See also the
discussion in Ref. \cite{koide}.

For the present case, these derivatives are given by 
\begin{eqnarray}
D\underline{\widehat{y}}(t) &=&\underline{u}(\underline{\widehat{y}}(t),t),
\\
\tilde{D}\underline{\widehat{y}}(t) &=&\underline{\tilde{u}}(\underline{%
\widehat{y}}(t),t).
\end{eqnarray}

Similarly to the correspondence principle in quantum mechanics, we require
that the action which obtained by substituting these new stochastic
variables should reduce to the corresponding classical one in the vanishing
noise limit, $\nu \rightarrow 0$. Moreover, the time reversal symmetry
should be satisfied. Thus, as is discussed in Refs. \cite{kk2,kk3,koide},
the time derivative term should be given by the average of the contributions
of the mean forward and backward derivatives \cite{yasue,kk2}. Then a
reasonable stochastic representation of Eq. (\ref{LatticeAction}) is given
by 
\begin{equation}
I_{SVM}[\widehat{y}]=\int_{t_{i}}^{t_{f}}dt\ E\left[ \frac{\sigma }{4}%
\left\{ (D\underline{\widehat{y}}\ast D\underline{\widehat{y}})_{\Delta x}+(%
\tilde{D}\underline{\widehat{y}}\ast \tilde{D}\underline{\widehat{y}}%
)_{\Delta x}\right\} +\frac{T}{2}(\underline{\widehat{y}}\ast \Delta _{x}%
\underline{\widehat{y}})_{\Delta x}\right] .  \label{string_SLA}
\end{equation}

The stochastic variation is applied to this stochastic action in the
following manner. We consider the following variation of the stochastic
variable, 
\begin{equation}
\widehat{y}_{n}(t)\longrightarrow \widehat{y}_{n}(t)+\eta (\widehat{y}%
_{n}(t),t),
\end{equation}%
where $\eta (x,t)$ is an arbitrary, but infinitesimal smooth function of $x$
and $t,$ satisfying the boundary condition, 
\begin{equation}
\eta (x,t_{i(f)})=0.
\end{equation}%
The detailed operation of the stochastic variation is the same as that in
the particle systems. See Ref. \cite{kk2} for details. Finally, we obtain
the following result, 
\begin{equation}
\left\{ \partial _{t}+\underline{v}(\underline{y},t)\cdot \underline{\nabla }%
_{y}\right\} \underline{v}(\underline{y},t)-\frac{2\nu ^{2}}{(\Delta x)^{2}}%
\underline{\nabla }_{y}\left\{ \frac{1}{\sqrt{\rho }}\nabla _{y}^{2}\sqrt{%
\rho }\right\} =\frac{T}{\sigma }\Delta _{x}\underline{y}(t),  \label{EOM}
\end{equation}%
where $\nabla _{y}^{2}=\underline{\nabla }_{y}\cdot \underline{\nabla }_{y}$%
. Here we introduced the mean velocity $\underline{v}\left( \underline{y}%
,t\right) $ as a function of $\underline{y}$ by 
\begin{equation}
\underline{v}(\underline{y},t)=\frac{\underline{u}(\underline{y},t)+%
\underline{\tilde{u}}(\underline{y},t)}{2}.  \label{st-euler}
\end{equation}%
By using this, the two Fokker-Planck equations introduced before are
expressed by the following unified form, 
\begin{equation}
\partial _{t}\rho (\underline{y},t)=-\underline{\nabla }_{y}\cdot (\rho (%
\underline{y},t)\underline{v}(\underline{y},t)).  \label{st-fp}
\end{equation}%
This is nothing but the equation of continuity. The two equations (\ref{EOM}%
) and (\ref{st-fp}) are the final result of SVM.

It should be mentioned that Eq. (\ref{EOM}) is a kind of the Euler equation
(ideal fluid equation) in the space of the amplitude configuration. The
corresponding equation in quantum mechanics of a particle system is known in
the Mandelung-Bohm theory \cite{man,bohm,taka,wall,holland,wyatt,durr}. In
Eq. (\ref{EOM}), the quantum effect is induced only by the last term on the
left hand side and this term is the generalization of the so-called quantum
potential.

These results can be simplified by introducing the wave function for the
amplitude configuration as 
\begin{equation}
\psi (\underline{y},t)=\sqrt{\rho (\underline{y},t)}e^{i\theta (\underline{y}%
,t)},  \label{Sch_functional_y}
\end{equation}%
where the phase is defined by 
\begin{equation}
\underline{v}(\underline{y},t)=\frac{2\nu }{\Delta x}\underline{\nabla }%
_{y}\theta (\underline{y},t).
\end{equation}

In this paper, anticipating the continuum limit of our SVM results, we refer
to Eq.(\ref{Sch_functional_y}) as \textit{wave functional}. Then all other
functions of $\underline{y},$ such as $\underline{v}$ and $\rho $ become
also corresponding functionals.

From Eqs. (\ref{EOM}) and (\ref{st-fp}), the equation for the wave
functional is given by 
\begin{equation}
i \partial _{t}\psi (\underline{y},t)= \tilde{H} \psi (\underline{y},t),
\label{Schger_y}
\end{equation}%
where 
\begin{equation}
\tilde{H} = \Delta x\left[ -\frac{\nu }{(\Delta x)^{2}}\nabla_{y}^{2} + 
\frac{1}{4\nu}\frac{T}{\sigma }\underline{y}^{T}\cdot \Delta _{x}\underline{y%
}\right] .  \label{Hamilton_string}
\end{equation}%
In the continuum limit, the derivative operators, $\underline{\nabla }_{y}$
and $\nabla _{y}^{2}$ can be expressed with the respective functional
derivatives. See the discussion in Sec. \ref{Sec:Cont} for details.

To complete the quantization, we need to specify the intensity of the noise $%
\nu$ and confirm that $\hbar \tilde{H}$ can be interpreted as the
Hamiltonian operator. Thus we cannot yet call this equation as the Schr\"{o}%
dinger equation of the amplitude configuration. In the usual SVM, this is
determined in a heuristic way, that is, it is fixed so as to reproduce the
Schr\"{o}dinger equation. Then, in the present case, we find 
\begin{equation}
\nu =\frac{\hbar }{2\sigma },
\end{equation}
because $\sigma$ is the mass of the string.

Later, in the quantization of the complex Klein-Gordon equation, we discuss
how this $\nu$ is determined from the direct experimental evidences such as
Planck-Einstein-de Broglie's relation, $E=\hbar \omega$ with $\omega$ being
a frequency, without assuming the Schr\"{o}dinger equation. See Sec. \ref%
{chap:nu} for details.

\subsection{SVM quantization: $k$-space}

The above SVM quantization on space lattice action (\ref{string_SLA}) should
be invariant in any choice of unitary equivalent variables. In particular,
if we choose the unitary transform which diagonalize the Laplacian matrix,
this should become equivalent to the Fourier transform of the field in the
continuum limit. Due to the presence of the periodic boundary condition,
this equivalence is not obvious. In this section we verify this explicitly. 

Let us denote $N$ eigenvectors of $\Delta _{x}$ as $\underline{u}^{\left(
n\right) }$, 
\begin{equation}
\Delta _{x}\underline{u}^{\left( n\right) }=-k_{n}^{2}\underline{u}^{\left(
n\right) },
\end{equation}%
where $n = 0, \cdots, N-1$ and 
\begin{equation}
k_{n}=\frac{2}{\Delta x}\sin \left( n\frac{\Delta k\Delta x}{2}\right) ,
\end{equation}%
with $\Delta k$ being $2\pi /L$. These eigenvectors are normalized as 
\begin{equation}
(\underline{u}^{\left( l\right) }\ast \underline{u}^{\left( m\right) })_{x}=%
\frac{1}{\Delta k}\delta _{lm}.  \label{orthogonality_of_u}
\end{equation}%
The explicit forms of these eigenvectors are shown in Appendix \ref{app}.

Any $N$-vector $\underline{f}$ in the $x$-space can be expanded with these
bases, so that we write the amplitude of the string $\underline{y}$ as 
\begin{equation}
\underline{y}=\Delta k\sum_{n=0}^{N-1}c_{n}\underline{u}^{\left( n\right) }.
\label{y=Cu_text}
\end{equation}%
The set of coefficients $c_{n}$ contains equivalent information to $%
\underline{y}$, and we also denote it in the $N$-vector form as 
\begin{equation}
\underline{c}=\left( 
\begin{array}{c}
c_{0} \\ 
\vdots \\ 
c_{N-1}%
\end{array}%
\right) .
\end{equation}%
Thus we can use $\underline{c}$ instead of $\underline{y}$ to represent the
string dynamics. We refer to this representation as the $k$-space
representation, because, as is shown in Appendix \ref{app}, the vector $%
\underline{c}$ converges to the set of the usual Fourier coefficients in the
continuum limit, if the original function $y\left( x\right) $ is smooth
almost everywhere.

Then we can diagonalize $\Delta _{x}$ with these eigenvectors defining the
matrix $O$ as 
\begin{equation}
O^{T}\Delta _{x}O=\frac{1}{\Delta k\Delta x}\Lambda ,
\end{equation}
where $O$ is given by 
\begin{equation}
O=(\underline{u}^{\left( 0\right) },\underline{u}^{\left( 1\right) },\cdots ,%
\underline{u}^{\left( N-1\right) }).
\end{equation}%
Here the diagonal matrix $\Lambda $ is given by $\Lambda
=diag(-k_{0}^{2},-k_{1}^{2},-k_{2}^{2},\cdots ,-k_{N-1}^{2})$.

Due to the orthogonalization with the measure introduced in the definition
of the scalar product in the $x$-space, we also need to introduce a measure
in the scalar product for the $k$-space vectors. Then, using Eq. (\ref%
{y=Cu_text}), the two $x$-space vectors $\underline{f}$ and $\underline{h}$
are expressed with the corresponding $k$-space vectors $\underline{c}$ and $%
\underline{b}$ as 
\begin{eqnarray}
\underline{f} &=&\Delta k\ O\ \underline{c}\ , \\
\underline{h} &=&\Delta k\ O\ \underline{b}\ ,
\end{eqnarray}%
respectively. Then the scalar products are invariant for the
\textquotedblleft orthogonal" transform by $O$, 
\begin{equation}
(\underline{f}\ast \underline{h})_{x}=(\underline{c}\ast \underline{b})_{k},
\label{x-k}
\end{equation}%
as far as the scalar product of the $k$-space vectors $\underline{c}$ and $%
\underline{b}$ is defined as 
\begin{equation}
\left( \underline{c}\ast \underline{b}\right) _{k}\equiv \Delta
k\sum_{n=0}^{N-1}c_{n}b_{n}.  \label{k-ScalarProduct}
\end{equation}%
See Appendix \ref{app} for details. The above results show how we move from
one representation to the other.

Using this basis, the $k$-space representation of the classical lattice
action given by Eq. (\ref{LatticeAction}) is represented as 
\begin{eqnarray}
I[c] &=&\int dt\left[ \frac{\sigma }{2}\left( \frac{\partial \underline{c}}{%
\partial t}\ast \frac{\partial \underline{c}}{\partial t}\right) _{\Delta k}+%
\frac{T}{2}(\underline{c}\ast \Lambda \underline{c})_{\Delta k}\right] 
\nonumber \\
&=&\int dt\ \Delta k\sum_{n=0}^{N-1}\left[ \frac{\sigma }{2}\left( \frac{%
\partial c_{n}}{\partial t}\right) ^{2}-\frac{T}{2}k_{n}^{2}c_{n}^{2}\right]
,  \label{LatticeActionFourier}
\end{eqnarray}

To implement the stochastic variation, we assume the following forward and
backward SDEs for $\underline{c}$ as 
\begin{eqnarray}
d\widehat{c}_{n}(t) &=&u_{n}(\underline{\widehat{c}}(t),t)dt+\sqrt{\frac{%
2\nu ^{\prime }}{\Delta k}}dW_{n}^{\prime }(t), \\
d\widehat{c}_{n}(t) &=&\tilde{u}_{n}(\underline{\widehat{c}}(t),t)dt+\sqrt{%
\frac{2\nu ^{\prime }}{\Delta k}}d\tilde{W}_{n}^{\prime }(t),
\end{eqnarray}%
respectively. Here the functional velocity fields $\underline{u}(\underline{c%
},t)$ and $\underline{\tilde{u}}(\underline{c},t)$ are smooth functions of $(%
\underline{c},t)$, and $\underline{W^{\prime }}(t)$ and $\underline{\tilde{W}%
^{\prime }}(t)$ are again two independent Wiener processes with the
following correlation properties, 
\begin{eqnarray}
E\left[ d\underline{W^{\prime }}(t)\right] &=&E\left[ d\underline{\tilde{W}%
^{\prime }}\right] =0, \\
\hspace{-1cm}E\left[ d\underline{W^{\prime }}(t)d\underline{W^{\prime }}%
^{T}(t)\right] &=&E\left[ d\underline{\tilde{W}^{\prime }}d\underline{\tilde{%
W}^{\prime }}^{T}\right] =|dt|\ I.
\end{eqnarray}

The distribution function of the $\underline{c}$ configuration is defined by 
\begin{equation}
\rho (\underline{c},t)=E[\delta ^{\left( N\right) }(\underline{c}-\underline{%
\widehat{c}}(t))].
\end{equation}%
Then the self-consistency condition is given by 
\begin{equation}
\underline{u}(\underline{c},t)=\underline{\tilde{u}}(\underline{c},t)+\frac{%
2\nu ^{\prime }}{\Delta k}\underline{\nabla }_{c}\ln \rho (\underline{c},t),
\end{equation}
and the Fokker-Planck equation is expressed as 
\begin{equation}
\partial _{t}\rho (\underline{c},t)=-\underline{\nabla }_{c}\cdot \left\{
\rho (\underline{c},t)\underline{v}(\underline{c},t)\right\}.
\end{equation}%
Here $\underline{\nabla }_{c}=(\partial /\partial c_{0},\cdots ,\partial
/\partial c_{N-1})$ and the mean functional velocity $\underline{v} $ is
defined by 
\begin{equation}
\underline{v}(\underline{c},t)=\frac{\underline{u}(\underline{c},t)+%
\underline{\tilde{u}}(\underline{c},t)}{2}.
\end{equation}

The intensity of the noise $\nu ^{\prime }$ can be calculated in two
different ways. One is to compare to the SDEs in the $x$-space by using $%
\underline{\hat{y}}(t)=(\Delta k)O\underline{\hat{c}}(t)$. The other to
employ the method which is introduced in Sec. \ref{chap:nu}\textbf{.} We
obtain the same result in either case and then the intensity is given by 
\begin{equation}
\nu ^{\prime }=\nu =\frac{\hbar }{2\sigma }.
\end{equation}

By applying the stochastic variation to the stochastic action (\ref%
{LatticeActionFourier}), we obtain 
\begin{equation}
\left\{ \partial _{t}+\underline{v}(\underline{c},t)\cdot \underline{\nabla }%
_{c}\right\} \ \underline{v}(\underline{c},t)-\frac{\hbar ^{2}}{2\sigma
^{2}(\Delta k)^{2}}\underline{\nabla }_{c}\left\{ \frac{1}{\sqrt{\rho }}%
\nabla _{c}^{2}\sqrt{\rho }\right\} =\frac{T}{\sigma }\Lambda \underline{y}%
(t).
\end{equation}

Similarly to the case of the $x$-space, the equation for the wave functional
corresponding to the above dynamics can be introduced as 
\begin{equation}
i \partial _{t}\psi (\underline{c},t) = \tilde{H} \psi (\underline{c},t),
\label{Schger_c}
\end{equation}%
where 
\begin{equation}
\tilde{H} =\left[ -\frac{\hbar }{2\sigma (\Delta k)}\nabla _{c}^{2}+\Delta k%
\frac{\sigma \omega _{n}^{2}\ }{2\hbar}\underline{c}^{2}\right] ,
\label{Hamilton_c}
\end{equation}%
with $\omega _{n}^{2}=k_{n}^{2}T/\sigma $. Here, the wave functional in the $%
k$-space is given by 
\begin{equation}
\psi (\underline{c},t)=\sqrt{\rho (\underline{c},t)}e^{i\theta (\underline{c}%
,t)},
\end{equation}%
and the phase is defined with the mean functional velocity as 
\begin{equation}
\underline{v}(\underline{c},t)=\frac{\hbar }{\sigma \ \Delta k}\underline{%
\nabla }_{c}\theta (\underline{c},t).
\end{equation}

\section{Complex Klein-Gordon field}

In this section, we discuss the SVM quantization of the complex Klein-Gordon
equation by using the definitions introduced in the previous section. The
ambiguity for the introduction of the wave functional is also discussed.

\subsection{SVM quantization on lattice}

The classical Lagrangian of the complex Klein-Gordon equation is 
\begin{equation}
L=\int_{V}d^{3}\mathbf{x}\left[ \frac{1}{c^{2}}(\partial _{t}\phi ^{\ast }(%
\mathbf{x},t))(\partial _{t}\phi (\mathbf{x},t))-\nabla \phi ^{\ast }(%
\mathbf{x},t)\cdot \nabla \phi (\mathbf{x},t)-\mu ^{2}\phi ^{\ast }(\mathbf{x%
},t)\phi (\mathbf{x},t)\right] ,  \label{LocalLag}
\end{equation}%
where $V=L^{3}$ is the space volume and $\mu =mc/\hbar $ with $m$ and $c$
being the mass and the speed of light, respectively. We apply the periodic
boundary condition at the space boundary.

To implement the stochastic variation, we need to re-define the system on a
spatial lattice as discussed in the previous section, and consider the
stochastic dynamics on each grid. Let us express the filed on the grid
points as $\phi_{\mathbf{x}}$. Then the stochastic Lagrangian is expressed
as 
\begin{eqnarray}
L_{SVM} &=& E\left[ \frac{1}{2c^{2}}\left\{ (D\underline{\widehat{\phi }}%
^{\ast }\ast D\underline{\widehat{\phi }})_{\mathbf{x}}+(\tilde{D}\underline{%
\widehat{\phi }}^{\ast }\ast \tilde{D}\underline{\widehat{\phi }})_{\mathbf{x%
}}\right\} \right.  \nonumber \\
&&\left. +(\underline{\widehat{\phi }}^{\ast }\ast \mathbf{\Delta }_{\mathbf{%
x}}\underline{\widehat{\phi }})_{\mathbf{x}}-\mu ^{2}(\underline{\widehat{%
\phi }}^{\ast }\ast \underline{\widehat{\phi }})_{\mathbf{x}}\right] ,
\label{lag_x_1}
\end{eqnarray}%
where $\mathbf{x}$ indicates the grid points, and we introduced the vector
notation $\underline{\phi }=$ $\left\{ \phi_{\mathbf{x}} \right\} $ with the
scalar product extended in the three dimensional $\mathbf{x}$-space as 
\begin{equation}
\left( \underline{f}\ast \underline{h}\right) _{\mathbf{x}}\equiv \Delta ^{3}%
\mathbf{x}\sum_{\mathbf{x}}f_{\mathbf{x}}h_{\mathbf{x}},
\end{equation}%
where $\Delta ^{3}\mathbf{x}$ is the volume of unit lattice cube, $(\Delta
x)^3$, and $\sum_{\mathbf{x}}$ indicates the sum over all lattice grids. The
Laplacian matrix $\mathbf{\Delta }_{\mathbf{x}}$ is also extended to the
three dimensional $\mathbf{x}$-space. See Appendix \ref{app} for details.

As was done in the classical string, it is possible to implement SVM
directly in the $\mathbf{x}$-space. We will, however, discuss the relation
between SVM and the usual canonical quantization and, for this purpose,
should develop the quantization in the $\mathbf{k}$-space. The result in the 
$\mathbf{x}$-space is given in Appendix \ref{app:x}.

By using the orthogonal transform which diagonalizes the matrix $\mathbf{%
\Delta }_{\mathbf{x}}$, the field configuration defined at the each grid
point $\mathbf{x}$ is transformed to the corresponding amplitude in the $%
\mathbf{k}$-space $\widehat{C}_{\mathbf{k}}(t)$ as 
\begin{equation}
\widehat{\phi }_{\mathbf{x}}(t)=\sqrt{\Delta ^{3}\mathbf{k}}\sum_{\mathbf{k}}%
\widehat{C}_{\mathbf{k}}(t)\frac{e^{i\mathbf{kx}}}{\sqrt{V}},
\label{FourierSt}
\end{equation}%
where $\Delta ^{3}\mathbf{k}=(\Delta k)^3=\left( 2\pi \right) ^{3}/V$, and $%
\widehat{C}_{\mathbf{k}}(t)$ is the stochastic complex variable written with
two real stochastic variables, $\widehat{C}_{R,\mathbf{k}}(t)$ and $\widehat{%
C}_{I,\mathbf{k}}(t)$ as%
\begin{equation}
\widehat{C}_{\mathbf{k}}(t)=\frac{\widehat{C}_{R,\mathbf{k}}(t)+i\widehat{C}%
_{I,\mathbf{k}}(t)}{\sqrt{2}}.
\end{equation}

It should be emphasized once again that the one-to-one correspondence
between $\hat{\phi}$ and $\hat{C}$ expressed by Eq.(\ref{FourierSt}) is just
an \emph{analog} of the Fourier transform defined between the two discrete
sets of variables. It coincides with the Fourier transform when and only
when these variables are smooth and the continuum limit is taken. See
Appendix \ref{app} for details.

With this, the stochastic action is given by 
\begin{eqnarray}
\lefteqn{I_{SVM}[\underline{\widehat{C}}_{R},\underline{\widehat{C}}_{I}]} 
\nonumber \\
&=&\frac{1}{2}\int_{t_{i}}^{t_{f}}dt\sum_{i=R,I}E\left[ \frac{1}{2c^{2}}%
\left\{ (D\underline{\widehat{C}}_{i}\ast D\underline{\widehat{C}}_{i})_{%
\mathbf{k}}+(\tilde{D}\underline{\widehat{C}}_{i}\ast \tilde{D}\underline{%
\widehat{C}}_{i})_{\mathbf{k}}\right\} -(\underline{\widehat{C}}_{i}\ast
\Lambda \underline{\widehat{C}}_{i})_{\mathbf{k}}\right]  \nonumber \\
&=&\frac{1}{2}\int_{t_{i}}^{t_{f}}dt\sum_{i=R,I}\sum_{\mathbf{k}}\Delta ^{3}%
\mathbf{k}  \nonumber \\
&&\times E\left[ \frac{1}{2c^{2}}\left\{ (D\widehat{C}_{i,\mathbf{k}}(t))(D%
\widehat{C}_{i,\mathbf{k}}(t))+(\tilde{D}\widehat{C}_{i,\mathbf{k}}(t))(%
\tilde{D}\widehat{C}_{i,\mathbf{k}}(t))\right\} -(\mathbf{q}_{\mathbf{k}%
}^{2}+\mu ^{2})\widehat{C}_{i,\mathbf{k}}^{2}(t)\right] ,
\end{eqnarray}%
where 
\begin{equation}
\Lambda _{\mathbf{k,k^{\prime }}}=(\mathbf{q}_{\mathbf{k}}^{2}+\mu
^{2})\delta _{\mathbf{k,k^{\prime }}}  \label{lambda_k}
\end{equation}
with 
\begin{equation}
\mathbf{q}_{\mathbf{k}}=\frac{2}{\Delta x}\left( 
\begin{array}{c}
\sin \left( \Delta x\ k_{x}/2\right) \\ 
\sin \left( \Delta x\ k_{y}/2\right) \\ 
\sin \left( \Delta x\ k_{z}/2\right)%
\end{array}%
\right) .  \label{q_k}
\end{equation}%
Note that this eigenvalue $\mathbf{q}_{\mathbf{k}}^{2}$ reduces simply to $%
\mathbf{k}^{2}$ for $\left\vert (\Delta x)\mathbf{k}\right\vert \ll 1.$

The forward and backward SDEs for these stochastic variables, $\widehat{C}%
_{R,\mathbf{k}}(t)$ and $\widehat{C}_{I,\mathbf{k}}(t)$ are, respectively,
given by 
\begin{eqnarray}
d\widehat{C}_{i,\mathbf{k}}(t) &=&u_{i,\mathbf{k}}( \underline{\widehat{C}}%
_{R}(t),\underline{\widehat{C}}_{I}(t),t)dt+\sqrt{\frac{2\nu }{\Delta ^{3}%
\mathbf{k}}}d{W}_{i,\mathbf{k}}(t), \\
d\widehat{C}_{i,\mathbf{k}}(t) &=&\tilde{u}_{i,\mathbf{k}}(\underline{%
\widehat{C}}_{R}(t),\underline{\widehat{C}}_{I}(t),t)dt+\sqrt{\frac{2\nu }{%
\Delta ^{3}\mathbf{k}}}d\tilde{W}_{i,\mathbf{k}}(t),\ 
\end{eqnarray}%
and the correlation properties of the noises are 
\begin{eqnarray}
E\left[ dW_{i,\mathbf{k}}(t)\right] &=&E\left[ d\tilde{W}_{i,\mathbf{k}}(t)%
\right] =0, \\
E\left[ dW_{j,\mathbf{k^{\prime }}}(t)dW_{i,\mathbf{k}}(t)\right] &=&E\left[
d\tilde{W}_{j,\mathbf{k^{\prime }}}(t)d\tilde{W}_{i,\mathbf{k}}(t)\right]
=\delta _{jk}\delta _{\mathbf{k},\mathbf{k^{\prime }}}^{(3)}|dt|,
\end{eqnarray}%
Here the indices $i,j$ denote $R$ or $I$. As before, there is no correlation
between $dW_{i,\mathbf{k}}(t)$ and $d\tilde{W}_{i,\mathbf{k}}(t)$.
Note that, differently from the SVM application to particle systems, 
the above Wiener process has a correlation for ${\bf k}$ which is proportional to the 
Dirac delta function in the continuum limit. 
Thus the existence of the well-defined continuum limit should be verified.
This is discussed in Sec. \ref{Sec:Cont}.

The distribution for the $\mathbf{k}$-space configuration $(\underline{{C}}%
_{R}(t),\underline{{C}}_{I}(t))$ is then defined by 
\begin{equation}
\rho (\underline{{C}}_{R},\underline{{C}}_{I},t)=E\left[ \prod_{\mathbf{k}%
}\delta (C_{R,\mathbf{k}}-\widehat{C}_{R,\mathbf{k}}(t))\delta (C_{I,\mathbf{%
k}}-\widehat{C}_{I,\mathbf{k}}(t))\right] .
\end{equation}%
The self-consistency condition is derived from the equivalence of the two
Fokker-Planck equations of the two SDEs introduced above as 
\begin{equation}
u_{i,\mathbf{k}}(\underline{{C}}_{R},\underline{{C}}_{I},t)=\tilde{u}_{i,%
\mathbf{k}}(\underline{{C}}_{R},\underline{{C}}_{I},t)+\frac{2\nu }{\Delta
^{3}\mathbf{k}}\frac{\partial }{\partial {C_{i,\mathbf{k}}}}\ln \rho (%
\underline{{C}}_{R},\underline{{C}}_{I},t).
\end{equation}%
Using this condition the Fokker-Planck equation is expressed as 
\begin{equation}
\partial _{t}\rho (\underline{{C}}_{R},\underline{{C}}_{I},t)=-\sum_{j=I,R}%
\sum_{\mathbf{k}}\frac{\partial }{\partial C_{j,\mathbf{k}}}\{\rho (%
\underline{{C}}_{R},\underline{{C}}_{I},t)v_{j,\mathbf{k}}(\underline{{C}}%
_{R},\underline{{C}}_{I},t)\}.  \label{FK1}
\end{equation}%
Here we used the mean functional velocity defined by 
\begin{equation}
v_{i,\mathbf{k}}(\underline{{C}}_{R},\underline{{C}}_{I},t)=\frac{u_{i,%
\mathbf{k}}(\underline{{C}}_{R},\underline{{C}}_{I},t)+\tilde{u}_{i,\mathbf{k%
}}(\underline{{C}}_{R},\underline{{C}}_{I},t)}{2}.
\end{equation}

With these definitions, the stochastic variation leads to the following
functional Euler equation for the complex Klein-Gordon equation, 
\begin{eqnarray}
&&\left( \partial _{t}+\sum_{j=R,I}\sum_{\mathbf{k^{\prime }}}v_{j,\mathbf{%
k^{\prime }}}(\underline{{C}}_{R},\underline{{C}}_{I},t)\frac{\partial }{%
\partial C_{j,\mathbf{k^{\prime }}}}\right) v_{i,\mathbf{k}}(\underline{{C}}%
_{R},\underline{{C}}_{I},t)  \nonumber \\
&&-\frac{2\nu ^{2}}{(\Delta ^{3}\mathbf{k})^{2}}\frac{\partial }{\partial
C_{i,\mathbf{k}}}\left\{ \rho ^{-1/2}(\underline{{C}}_{R},\underline{{C}}%
_{I},t)\sum_{j=R,I}\sum_{\mathbf{k^{\prime }}}\left( \frac{\partial }{%
\partial C_{j,\mathbf{k^{\prime }}}}\right) ^{2}\rho ^{1/2}(\underline{{C}}%
_{R},\underline{{C}}_{I},t)\right\}  \nonumber \\
&=&-\frac{1}{2}\frac{\partial }{\partial C_{i,\mathbf{k}}}\sum_{j=R,I}\sum_{%
\mathbf{k^{\prime }}}\omega _{\mathbf{k}^{\prime }}^{2}C_{j,\mathbf{%
k^{\prime }}}^{2},  \label{eq-v-kg1}
\end{eqnarray}%
where 
\begin{equation}
\omega _{\mathbf{k}}=c\sqrt{\mathbf{q}_{\mathbf{k}}^{2}+\mu ^{2}}.
\end{equation}%
The two equations (\ref{FK1}) and (\ref{eq-v-kg1}) compose a closed system
for our unknown variables $\rho $, $v_{R,\mathbf{k}}$ and $v_{I,\mathbf{k}}$
which determine completely the dynamics of the system in terms of the field
configuration.

We now introduce the wave functional for the field configuration in the $%
\mathbf{k}$-space as 
\begin{equation}
\psi (\underline{{C}}_{R},\underline{{C}}_{I},t)=\rho (\underline{{C}}_{R},%
\underline{{C}}_{I},t)e^{i\theta (\underline{{C}}_{R},\underline{{C}}%
_{I},t)},  \label{Psi}
\end{equation}%
where the phase is defined by 
\begin{equation}
v_{i,\mathbf{k}}(\underline{{C}}_{R},\underline{{C}}_{I},t)=\frac{2\nu }{%
\Delta ^{3}\mathbf{k}}\frac{\partial }{\partial C_{i,\mathbf{k}}}\theta (%
\underline{{C}}_{R},\underline{{C}}_{I},t).  \label{phase-kk}
\end{equation}

The dynamics of this wave functional is determined by the following
functional equation,%
\begin{equation}
i\partial _{t}\psi (\underline{{C}}_{R},\underline{{C}}_{I},t)=\tilde{H}\psi
(\underline{{C}}_{R},\underline{{C}}_{I},t),  \label{sch_eq_am}
\end{equation}%
with the time evolution operator 
\begin{equation}
\tilde{H}=\Delta ^{3}\mathbf{k}\sum_{\mathbf{k}}\sum_{j=R,I}\left[ -\frac{%
\nu }{(\Delta ^{3}\mathbf{k})^{2}}\left( \frac{\partial }{\partial C_{j,%
\mathbf{k}}}\right) ^{2}+\frac{1}{4\nu }\omega _{\mathbf{k}}^{2}C_{j,\mathbf{%
k}}^{2}\right] .  \label{Ham_am}
\end{equation}

As was already mentioned in the discussion of the string, to complete the
procedure of the quantization, we need to specify the value of the noise
intensity and confirm that the time-evolution operator $\tilde{H}$ is
determined by the Hamiltonian operator.

\subsection{Ambiguity for definition of phase and divergent vacuum energy}

\label{chap:ambi}

For the definition of the phase, there is an ambiguity which has not yet
been discussed so far. Because of the definition of the phase in Eq. (\ref%
{phase-kk}), it is possible to add an arbitrary function $f(t)$, which is
independent of the field $C_{i,\mathbf{k}}$, in the evolution equation of
the phase as 
\begin{eqnarray}
\partial _{t}\theta  &=&-\sum_{i=I,R}\sum_{\mathbf{k}}\frac{\nu }{\Delta ^{3}%
\mathbf{k}}\left[ \left( \frac{\partial \theta }{\partial C_{i,\mathbf{k}}}%
\right) ^{2}-\rho ^{-1/2}\frac{\partial ^{2}}{\partial C_{i,\mathbf{k}}^{2}}%
\rho ^{1/2}\right] -\frac{\Delta ^{3}\mathbf{k}}{4\nu }\sum_{i=I,R}\sum_{%
\mathbf{k}}\omega _{\mathbf{k}}^{2}C_{i,\mathbf{k}}^{2}+f(t).  \nonumber \\
&&
\end{eqnarray}%
By using this ambiguity, we can eliminate the divergent vacuum energy. In
other word, this divergence is attributed to the artificial introduction of
the phase and does not appear in solving the coupled equation of the
Fokker-Planck equation and the functional Euler equation directly. See also
the argument around Eq. (\ref{vac}). It should be however, mentioned that
this procedure is not applicable if the vacuum state is not spatially
homogeneous as is the case of the Casimir effect because $f(t)$ is a
function only of time. 

The same ambiguity appears even in quantum mechanics. See the discussion
around Eq. (45) in Ref. \cite{kk2}.

\section{Determination of noise intensity}

In this section, we discuss the intensity of the noise $\nu$. We determine
this value so as to reproduce the single-particle energy. To introduce this
energy, the Fock state vector should be defined in this framework.

\subsection{Fock state vector as stationary solution}

\label{chap:wf}

In SVM, the dynamics of the quantized fields are described by Eq. (\ref%
{sch_eq_am}). This is the functional differential equation and the
stationary solution can be obtained explicitly. As will be shown soon later,
this stationary solution corresponds to the state vector in the Fock space.

The explicit form of this wave functional is obtained in the same manner as
the calculation of the harmonic oscillator potential in quantum mechanics.
For the free field, the dynamics of each $\mathbf{k}$ component is separated
and hence we introduce the following factorized ansatz for the stationary
state, 
\begin{equation}
\psi (\underline{{C}}_{R},\underline{{C}}_{I},t)=e^{-i\Omega \
t}\prod\limits_{\mathbf{k}}\psi _{\mathbf{k},R}(C_{R,\mathbf{k}})\psi _{%
\mathbf{k},I}(C_{I,\mathbf{k}}).  \label{decom}
\end{equation}%
Then the stationary eigenvalue equation becomes one unique form for all the
factorization index,$\ \alpha =\left( \mathbf{k},i=R,I\right) $ as 
\begin{equation}
\partial _{\xi }^{2}\psi _{\alpha }(\xi )+(\epsilon _{\alpha }-\xi ^{2})\psi
_{\alpha }(\xi )=0,  \label{epsilon}
\end{equation}%
where 
\begin{equation}
\xi =C_{i,\mathbf{k}}\sqrt{\frac{\omega _{k}}{2\nu }\ \Delta ^{3}\mathbf{k\ }%
}.  \label{gzi}
\end{equation}%
And the set of eigenvalues $\left\{ \epsilon _{\alpha }\right\} $ gives the
frequency $\Omega $ of the stationary state as%
\begin{equation}
\Omega =\frac{1}{2}\sum_{\alpha =\left( \mathbf{k,}i\right) }\epsilon
_{\alpha }\omega _{\mathbf{k}}.  \label{Omega}
\end{equation}

The solution of this differential equation is, as is well-known, expressed
with the Hermite polynomial, $H_{n}\left( \xi \right) $ as 
\begin{equation}
\Phi _{n}(\xi )=H_{n}(\xi )e^{-\xi ^{2}/2}.
\end{equation}%
In this case, $\epsilon _{\alpha }$ in Eq.(\ref{epsilon}) is given by%
\begin{equation}
\epsilon _{\alpha }=2n_{\alpha }+1,  \label{EigenValues}
\end{equation}
where $n_{\alpha }=0,\cdots, \infty$.

Finally, the stationary solution or the eigenstate for the component $%
\mathbf{k}$ is given by \footnote{%
It is possible to introduce relative phases in this expression.} 
\begin{eqnarray}
\psi _{\mathbf{k}}(C_{R,\mathbf{k}},C_{I,\mathbf{k}},t) &=&N_{\mathbf{k}%
}e^{-i\omega _{\mathbf{k}}(n+1)t}\sum_{m=0}^{n}H_{n-m}(C_{R,\mathbf{k}}\sqrt{%
\omega _{\mathbf{k}}/2\nu })H_{m}(C_{I,\mathbf{k}}\sqrt{\omega _{\mathbf{k}%
}/2\nu })  \nonumber \\
&&\times e^{- (\Delta^3 \mathbf{k}) \omega _{\mathbf{k}}(C_{R,\mathbf{k}%
}^{2}+C_{I,\mathbf{k}}^{2})/(4\nu) },  \label{eigenstate}
\end{eqnarray}%
with $N_{\mathbf{k}}$ being the normalization factor which is given by
initial conditions. In the above, we fixed the total frequency as 
\begin{equation}
\epsilon _{\mathbf{k}}=\epsilon _{\left( \mathbf{k},R\right) }+\epsilon
_{\left( \mathbf{k},I\right) } = \omega _{\mathbf{k}}(n+1).
\end{equation}%
The summation over $m$ is related to the two degrees of freedom, $i=R,I$.

In particular, the wave functional of the lowest eigenstate corresponds to $%
n_{\alpha }=0$ for all $\alpha ,$ and is given by 
\begin{equation}
\psi _{0}(\underline{{C}}_{R},\underline{{C}}_{I}) =\prod\limits_{\mathbf{k}%
}\left( \frac{(\Delta^3 \mathbf{k})\omega _{\mathbf{k}}}{2\pi \nu }%
\right)^{1/2} e^{-(\Delta^3 \mathbf{k})\omega _{\mathbf{k}}(C_{R,\mathbf{k}%
}^{2} + C_{I,\mathbf{k}}^{2})/(4\nu) }.  \label{vac}
\end{equation}%
Here we dropped the phase factor $e^{-i\omega _{\mathbf{k}}t}$ by using the
ambiguity $f(t)$ which appears in the definition of the phase as was
discussed in Sec. \ref{chap:ambi}. As for the normalization, we adopted 
\begin{equation}
\int \mathcal{D}\left[ C_{R}\right] \mathcal{D}\left[ C_{I}\right] |\psi
_{0}(\underline{{C}}_{R},\underline{{C}}_{I})|^{2}=1.
\end{equation}

The creation-annihilation operators are defined by 
\begin{eqnarray}
a_{\mathbf{k}}+b_{\mathbf{-k}} &=&A_{\mathbf{k}}C_{R,\mathbf{k}}+B_{\mathbf{k%
}}\frac{\partial }{\partial C_{R,\mathbf{k}}}, \\
a_{\mathbf{k}}^{\dagger }+b_{\mathbf{-k}}^{\dagger } &=&A_{\mathbf{k}}C_{R,%
\mathbf{k}}-B_{\mathbf{k}}\frac{\partial }{\partial C_{R,\mathbf{k}}}, \\
a_{\mathbf{k}}-b_{\mathbf{-k}} &=&i\left( A_{\mathbf{k}}C_{I,\mathbf{k}}+B_{%
\mathbf{k}}\frac{\partial }{\partial C_{I,\mathbf{k}}}\right) , \\
a_{\mathbf{k}}^{\dagger }-b_{\mathbf{-k}}^{\dagger } &=&-i\left( A_{\mathbf{k%
}}C_{I,\mathbf{k}}-B_{\mathbf{k}}\frac{\partial }{\partial C_{I,\mathbf{k}}}%
\right) ,
\end{eqnarray}%
where 
\begin{equation}
A_{\mathbf{k}}=\sqrt{\frac{\omega _{\mathbf{k}}}{2\nu }\Delta ^{3}\mathbf{k}}%
,~~~~B_{\mathbf{k}}=\sqrt{\frac{2\nu }{\omega _{\mathbf{k}}}\frac{1}{\Delta
^{3}\mathbf{k}}}.
\end{equation}%
One can easily confirm that 
\begin{eqnarray}
[a_{\mathbf{k}}, a_{\mathbf{k^{\prime }}}^\dagger] = [b_{\mathbf{k}}, b_{%
\mathbf{k^{\prime }}}^\dagger] = \delta^{(3)}_{\mathbf{k,k^{\prime }}}.
\end{eqnarray}

Then the lowest eigenstate satisfies 
\begin{equation}
a_{\mathbf{k}}\psi _{0}=b_{\mathbf{k}}\psi _{0}=0,
\end{equation}%
for $\forall \mathbf{k}$. Moreover, as seen from Eqs. (\ref{Omega}) and (\ref%
{EigenValues}), all the eigenstates of the wave functional can be specified
by a set of integers $\left\{ n_{\alpha }\right\}$ so that we interpret $%
n_\alpha$ as the number of quanta with $\alpha$. As is well-known in the
harmonic oscillator case, these states are obtained from the lowest
eigenstate by applying the creation operators $a_{\mathbf{k}}^{\dagger }$
and $b_{\mathbf{k}}^{\dagger }$, which is related to the Rodrigues formula, 
\begin{equation}
(\xi -\partial _{\xi })^{n}e^{-\xi ^{2}/2}=H_{n}(\xi )e^{-\xi ^{2}/2}.
\end{equation}%
Thus the Fock space vector constructed from $\phi_0$ by operating the
creation-annihilation operators are equivalent to the eigenstate of Eq. (\ref%
{sch_eq_am}).

In the lowest eigenstate, the width of the distribution in the field
configuration is proportional to $\nu $. Thus, in the vanishing noise limit $%
\nu \rightarrow 0$, this has a strong peak around $\left\{ C_{R,\mathbf{k}%
},C_{I,\mathbf{k}}\right\} =\left\{ 0,0\right\} $ where the energy of the
system disappears. This corresponds to the classical limit as we will see in
below.

Additionally, the lowest eigenstate is not normalizable for the massless
case $\mu =0$ because $\omega _{\mathbf{k=0}}=0.$ Thus, to construct a
consistent description in terms of the Fock state vectors, this mode should
be excluded from the definition of the lowest eigenstate as 
\begin{equation}
\psi (\underline{{C}}_{R},\underline{{C}}_{I},t)=\prod_{\mathbf{k}\neq
0}\psi _{\mathbf{k}}(C_{R,\mathbf{k}},C_{I,\mathbf{k}},t).
\end{equation}

\subsection{Noise intensity and Hamiltonian as the time evolution operator}

\label{chap:nu}

By using the arguments developed so far, we can determine the noise
intensity.

First we identify the quanta introduced above with physical particles and
hence the lowest eigenstate as the vacuum state. Then we can argue the
single-particle energy.

As shown in Ref. \cite{kk3}, the Hamiltonian $\mathcal{H}$ in SVM is
obtained by the Legendre transform of the stochastic Lagrangian. In our
case, it is given by 
\begin{eqnarray}
\mathcal{H} &=&E\left[ \frac{1}{c^{2}}\frac{(D\underline{\hat{\phi}}^{\ast
}\ast D\underline{\hat{\phi}})_{\mathbf{x}}+(\tilde{D}\underline{\hat{\phi}}%
^{\ast }\ast \tilde{D}\underline{\hat{\phi}})_{\mathbf{x}}}{2}-(\underline{%
\hat{\phi}}^{\ast }\ast \Delta _{\mathbf{x}}\underline{\hat{\phi}})_{\mathbf{%
x}}+\mu ^{2}(\underline{\hat{\phi}}^{\ast }\ast \underline{\hat{\phi}})_{%
\mathbf{x}}\right]  \nonumber \\
&=& \frac{1}{2c^2}\int D[C_R] D[C_I] \rho \sum_{i=R,I} \sum_{\mathbf{k}} %
\left[ \frac{4\nu^2}{\Delta^3 \mathbf{k}} \left\{ \left( \frac{\partial
\theta}{\partial C_{i,\mathbf{k}}} \right)^2 + \left( \frac{\partial \ln 
\sqrt{\rho}}{\partial C_{i,\mathbf{k}}} \right)^2 \right\} + \omega^2_{%
\mathbf{k}} C^2_{i,\mathbf{k}} \right].  \nonumber  \label{Hamilton_3} \\
\end{eqnarray}
Note that the above result can be re-expressed as 
\begin{equation}
\mathcal{H} = \langle \psi (t)|H|\psi (t)\rangle ,  \label{Hamilton_2}
\end{equation}%
where the Hamiltonian operator $H$ is defined by 
\begin{equation}
H=\frac{1}{2c^{2}}\sum_{i=R,I}\sum_{\mathbf{k}}\left[ -\frac{4\nu ^{2}}{%
\Delta ^{3}\mathbf{k}}\frac{\partial^2}{\partial^2 C_{i,\mathbf{k}}} +\omega
_{\mathbf{k}}^{2}C_{i,\mathbf{k}}^{2} \right] , \label{Hamilton_6}
\end{equation}%
and we introduced the expectation value of a functional $A(\underline{{C}}%
_{R},\underline{{C}}_{I},t)$ by 
\begin{equation}
\langle \psi (t)|A|\psi (t)\rangle =\int \prod_{\mathbf{k}}dC_{R,\mathbf{k}%
}dC_{I,\mathbf{k}}\ \psi ^{\ast }(\underline{{C}}_{R},\underline{{C}}%
_{I},t)A(\underline{{C}}_{R},\underline{{C}}_{I},t)\psi (\underline{{C}}_{R},%
\underline{{C}}_{I},t).  \label{phi-exp}
\end{equation}%
It should be noted that the Hamiltonian operator defined here differs 
from the time evolution operator $\tilde{H}$ (\ref{Ham_am}) by a constant factor 
depending the noise intensity. 
The same expression (\ref{Hamilton_6}) can be obtained also as Noether's charge 
by employing the invariance of the stochastic
action for the time translation by using the stochastic Noether theorem. 
For particle systems, see Ref. \cite{misawa3}.

Let us consider a single-particle state with a small wavenumber $\mathbf{k}$
which is given by operating a single $a_{\mathbf{k}}^{\dagger }$ to the
vacuum, 
\begin{equation}
\psi (\underline{{C}}_{R},\underline{{C}}_{I},t)=e^{-i\omega _{\mathbf{k}}\
t}\frac{1}{\sqrt{\Delta ^{3}\mathbf{k\ }}}a_{\mathbf{k}}^{\dagger }\psi _{0},
\end{equation}%
where, from Eq. (\ref{lambda_k}), $\omega_{\mathbf{k}}$ reduces to $c\sqrt{%
\mathbf{k}^{2}+\mu ^{2}}$ at least for small $\mathbf{k}$. As for the factor 
$1/\sqrt{\Delta ^{3}\mathbf{k\ }}$, see the discussion in Sec. \ref{Sec:Cont}%
. Substituting this into Eq. (\ref{Hamilton_3}), or equivalently Eq. (\ref%
{Hamilton_2}), the energy is given by 
\begin{equation}
\mathcal{H}=\frac{2\nu \mathbf{\ }}{c}\sqrt{\mathbf{k}^{2}+\mu ^{2}}.
\end{equation}%
Here the contribution from the vacuum polarization is subtracted. On the
other hand, from the Planck-Einstein-de Broglie relation, the
single-particle energy is known as $\hbar c \sqrt{\mathbf{k}^2 + \mu^2}$. To
reproduce this result, the intensity of the noise should be given by 
\begin{equation}
\nu =\frac{\hbar c^{2}}{2}.  \label{nuhbar}
\end{equation}

It should be noted that, substituting this result into the time evolution
operator $\tilde{H}$, we can find that the following relation is finally
satisfied, 
\begin{equation}
H=\hbar \tilde{H}.  \label{HHtil}
\end{equation}%
That is, Eq. (\ref{sch_eq_am}) can be interpreted as the functional Schr\"{o}%
dinger equation and the Hamiltonian operator is the generator of the time
evolution of this dynamics.

This relation is satisfied for other quantizations which we have studied so
far and very general. Therefore, we will propound Eq. (\ref{HHtil}) as the
condition to determine the intensity $\nu$ in the SVM quantization, instead
of calculating the single-particle energies for each case.

Finally, the functional Schr\"{o}dinger equation is expressed as 
\begin{eqnarray}
i\hbar \partial _{t}\psi (\underline{{C}}_{R},\underline{{C}}_{I},t) &=&H\
\psi (\underline{{C}}_{R},\underline{{C}}_{I},t)  \nonumber \\
&=&\Delta ^{3}\mathbf{k}\sum_{\mathbf{k}}\sum_{j=R,I}\left[ -\frac{\hbar
^{2}c^{2}}{2(\Delta ^{3}\mathbf{k})^{2}}\left( \frac{\partial }{\partial
C_{j,\mathbf{k}}}\right) ^{2}+\frac{1}{2c^{2}}\omega _{\mathbf{k}}^{2}C_{j,%
\mathbf{k}}^{2}\right] \psi (\underline{{C}}_{R},\underline{{C}}_{I},t). 
\nonumber \\
&&  \label{fse}
\end{eqnarray}%
In addition, by using the creation-annihilation operators, the Hamiltonian
operator is expressed as 
\begin{equation}
H=\sum_{\mathbf{k}}\hbar \omega _{\mathbf{k}}(a_{\mathbf{k}}^{\dagger }a_{%
\mathbf{k}}+b_{\mathbf{k}}^{\dagger }b_{\mathbf{k}}+1).  \label{Ham_op}
\end{equation}

It is well-known that the same functional Schr\"{o}dinger equation (\ref{fse}%
) can be obtained from the canonical quantization by substituting the
momentum field with the corresponding functional derivative \cite%
{holland,sym,lu,kie,huang}, although the above equation is still
discretized. Therefore it is concluded that the SVM quantization can reproduce
the result of the canonical quantization appropriately. 
On the other hand, the advantage of SVM to the canonical quantization 
becomes clear in the calculation of the Noether charge. 
Moreover, we need to confirm the existence of the well-defined continuum limit. 
In the next section, we will discuss these points.

\section{Other aspects in SVM quantization}

In this section, we discuss other aspects of the SVM quantization and
briefly mention properties of the functional Schr\"{o}dinger equation.

\subsection{Noether charge}

From Eq. (\ref{Ham_op}), we can reasonable observe that the energy is
expressed by the \textit{summation} of the particle and anti-particle
contributions. However, this is not yet sufficient to conclude that the
contribution from the anti-particles is precisely included. To verify the
equivalence between SVM and the canonical quantization, it is necessary also
to check the Noether charge, which should be given by the \textit{subtraction%
} of the contributions from particles and anti-particles. The Noether
theorem in SVM is discussed for the case of particle systems in Ref. \cite%
{misawa3}. We show that this can successfully be generalized for the case of
fields.

The Lagrangian for the complex Klein-Gordon equation is symmetric for the
global phase transform, 
\begin{equation}
\widehat{\phi }\longrightarrow \widehat{\phi }e^{i\alpha }\approx (1+i\alpha
)\widehat{\phi }=\widehat{\phi }+\delta \widehat{\phi },
\end{equation}%
where $\alpha $ is a small number. Then the stochastic action is modified as 
\begin{equation}
\delta I_{sto}=\frac{1}{2c^{2}}\int_{t_{i}}^{t_{f}}dt \frac{d}{dt}E\left[
(\delta \underline{\widehat{\phi }}^{\ast }* D\underline{\widehat{\phi }})_{%
\mathbf{x}} + (\delta \underline{\widehat{\phi }}*D\underline{\widehat{\phi }%
}^{\ast })_{\mathbf{x}} +(\delta \underline{\widehat{\phi }}^{\ast } * 
\tilde{D}\underline{\widehat{\phi }})_{\mathbf{x}} +(\delta \underline{ 
\widehat{\phi }} * \tilde{D}\underline{\widehat{\phi }}^{\ast })_{\mathbf{x}}%
\right] .
\end{equation}
In this derivation, we used the stochastic partial integration formula \cite%
{zm}, 
\begin{equation}
E[(D\widehat{X}(t))\widehat{Y}(t)+\widehat{X}(t)\tilde{D}\widehat{Y}(t)]=%
\frac{d}{dt}E[\widehat{X}(t)\widehat{Y}(t)],
\end{equation}%
assuming that the expectation value $E[~~]$ vanishes at the infinite
distance.

Thus the Noether charge is expressed in SVM as 
\begin{equation}
Q=-\frac{ie}{2\hbar c^{2}} E\left[ \left(\left\{ D\underline{\widehat{\phi }}%
_{\mathbf{x}}^{\ast }+\tilde{D} \underline{\widehat{\phi }}_{\mathbf{x}%
}^{\ast }\right\} * \underline{\widehat{\phi }}_{\mathbf{x}} \right)_{%
\mathbf{x}} -\left( \left\{ D\underline{\widehat{\phi }}_{\mathbf{x}}+\tilde{%
D} \underline{\widehat{\phi }}_{\mathbf{x}} \right\} * \underline{\widehat{%
\phi }}_{\mathbf{x}}^{\ast } \right)_{\mathbf{x}}\right] .
\end{equation}%
Here we choose $\alpha =-e/\hbar $. This expectation value is expressed with
the wave functional introduced above as 
\begin{eqnarray}
Q &=& e\int \mathcal{D}\left[ {C}_{R}\right] \mathcal{D}\left[ {C}_{I}\right]
\rho (\underline{{C}}_{R},\underline{{C}}_{I},t)\sum_{\mathbf{k}}\left[ 
\frac{\partial \theta (\underline{{C}}_{R},\underline{{C}}_{I},t)}{\partial
C_{I,\mathbf{k}}}C_{R,\mathbf{k}}-\frac{\partial \theta (\underline{{C}}_{R},%
\underline{{C}}_{I},t)}{\partial C_{R,\mathbf{k}}}C_{I,\mathbf{k}}\right] . 
\nonumber \\
&&
\end{eqnarray}%
By using the creation-annihilation operators introduced before, the above
expression is further rewritten as 
\begin{equation}
Q=e\langle \psi (t)|\sum_{\mathbf{k}}(a_{\mathbf{k}}^{\dagger }a_{\mathbf{k}%
}-b_{\mathbf{-k}}^{\dagger }b_{\mathbf{-k}})|\psi (t)\rangle.  \label{q-aad}
\end{equation}%
This is equivalent to the result of the canonical quantization.

Thus the SVM quantization developed so far can deal in the anti-particle
degrees of freedom correctly. It should be stressed that we do not suffer from 
the order of variables, differently from the canonical
quantization where the normal ordering product 
is introduced to obtain Eq. (\ref{q-aad}).

\subsection{Propagator}

We have shown that the theory obtained from SVM corresponds to the Schr\"{o}%
dinger picture in quantum field theory. On the other hand, the propagator
plays an important role in the Heisenberg and interaction pictures. This can
be introduced as follows.

In SVM, the expectation value of a quantity $A(\underline{{C}}_{R},%
\underline{{C}}_{I})$ at time $t$ is calculated by 
\begin{equation}
\left\langle A\left( t\right) \right\rangle =\int \mathcal{D}\left[ C_{R} %
\right] \mathcal{D}\left[ C_{I}\right] \rho (\underline{{C}}_{R},\underline{{%
C}}_{I},t) A(\underline{{C}}_{R},\underline{{C}}_{I}).  \label{expexp2}
\end{equation}

It should be stressed that the quantity $A(\underline{{C}}_{R},\underline{{C}}%
_{I})$ appearing in the Schr\"{o}dinger picture in the SVM quantization is
always given by the functional of $(\underline{{C}}_{R},\underline{{C}}_{I})$
and not an operator.

On the other hand, the Heisenberg operator of $A$ is introduced by 
\begin{equation}
A^{\left( H\right) }(\underline{{C}}_{R},\underline{{C}}_{I};t)
=e^{iHt/\hbar }A(\underline{{C}}_{R},\underline{{C}}_{I})e^{-iHt/\hbar }.
\label{op_hei}
\end{equation}%
This quantity behaves an operator because of the derivative operators contained in $H$.
For example, when $A(\{C_{R},C_{I}\})=C_{R,\mathbf{k}}$, the corresponding
Heisenberg operator is 
\begin{equation}
C_{R,\mathbf{k}}^{\left( H\right) }(t)=\cos (\omega _{\mathbf{k}}t)C_{R,%
\mathbf{k}}-i\frac{\hbar c^{2}}{2\omega _{\mathbf{k}}}\sin (\omega _{\mathbf{%
k}}t)\frac{\partial }{\partial C_{R,\mathbf{k}}}.  \label{C_Heisen}
\end{equation}

By using this operator, the expectation value of a general observable $A$
given by Eq. (\ref{expexp2}) is, in the Heisenberg picture, expressed as 
\begin{equation}
\left\langle A\right\rangle =\int \mathcal{D}\left[ C_{R}\right] \mathcal{D}%
\left[ C_{I}\right] \psi ^{\ast }(\underline{{C}}_{R},\underline{{C}}%
_{I},0)A^{\left( H\right) }(\underline{{C}}_{R},\underline{{C}}_{I};t)\psi (%
\underline{{C}}_{R},\underline{{C}}_{I},0).
\end{equation}%
In this form, the SVM expression for a propagator is written as 
\begin{eqnarray}
\lefteqn{i\hbar c\Delta _{F}(t,\mathbf{x})=\langle \phi ^{(H)}(\mathbf{x}%
,t)\phi ^{\ast }(\mathbf{0})\rangle \theta (t)+\langle \phi ^{\ast }(\mathbf{%
0})\phi ^{(H)}(\mathbf{x},t)\rangle \theta (-t)}  \nonumber \\
&=&\int \mathcal{D}\left[ C_{R}\right] \mathcal{D}\left[ C_{I}\right] \psi
_{0}^{\ast }(\underline{{C}}_{R},\underline{{C}}_{I})[\phi ^{(H)}(\mathbf{x}%
,t)\phi ^{\ast }(\mathbf{0})\theta (t)+\phi ^{\ast }(\mathbf{0})\phi ^{(H)}(%
\mathbf{x},t)\theta (-t)]\psi _{0}(\underline{{C}}_{R},\underline{{C}}_{I}),
\nonumber \\
&&
\end{eqnarray}%
where $\theta $ is the Heaviside step function and $\psi _{0}$ is the vacuum
wave functional given by Eq. (\ref{vac}). Using the expression (\ref%
{C_Heisen}) and performing the Gaussian integrals, we obtain finally%
\begin{equation}
\Delta _{F}(t,\mathbf{x})=\Delta ^{3}\mathbf{k}\sum_{\mathbf{k}}\int \frac{%
dk^{0}}{(2\pi )^{4}}\frac{e^{ikx}}{(k^{0})^2 - \mathbf{q}^2_{\mathbf{k}}
-\mu ^{2}+i\epsilon },  \label{prop}
\end{equation}%
where $kx = k^\mu x_\mu = k^{0}ct-\mathbf{k}\cdot \mathbf{x}$. Taking the
continuum limit $\Delta x \rightarrow 0$ where the upper limit of the
momentum integration goes to $\infty$, $\mathbf{q}_{\mathbf{k}}$ reduces to $%
\mathbf{k}$ and the above expression converges to the well-known Lorentz
covariant result, 
\begin{equation}
\Delta _{F}(t,\mathbf{x}) = \int \frac{dk^{4}}{(2\pi )^{4}}\frac{e^{ikx}}{%
k^{2}-\mu ^{2}+i\epsilon }.
\end{equation}%
See Sec. \ref{Sec:Cont} for more details of the continuum limit.

\subsection{Imaginary mass}

\begin{figure}[t]
\includegraphics[scale=0.3]{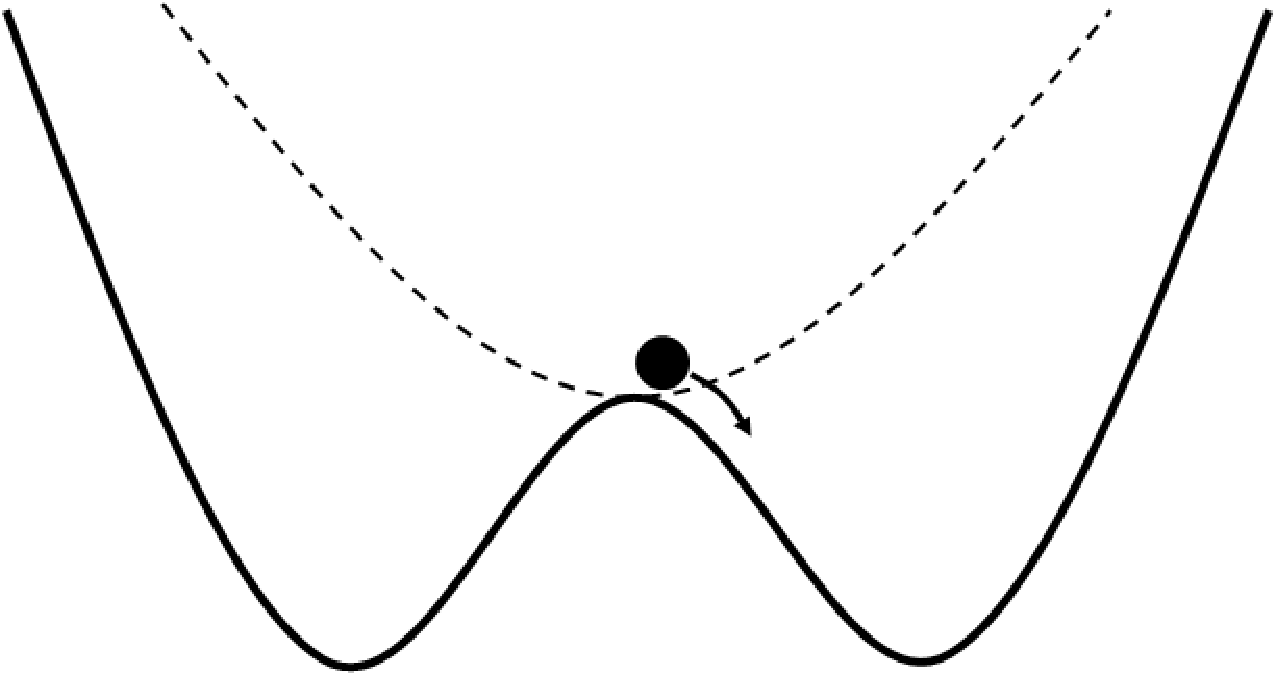}
\caption{The dynamics of the symmetry breaking. Due to the change of
external parameters, the initial potential denoted by the dashed line is
transformed to the solid line. }
\label{fig1}
\end{figure}

We have so far shown that the quantization in SVM reproduces the results of
the usual canonical quantization, when the Fock space representation is 
introduced. 
However we need not introduce the Fock space in the
procedure of the quantization, differently from the canonical quantization.
In principle, the quantum dynamics can be discussed by solving the the
functional Schr\"{o}dinger equation Eq. (\ref{fse}), or
equivalently the functional Euler equation Eq. (\ref{eq-v-kg1}), with Eq. (%
\ref{FK1}) directly with an in initial condition.

Furthermore, there exist situations where the construction of the Fock space
is not well-known. As an example of such a situation, let us consider the
Klein-Gordon equation with the imaginary mass, $\mu =i\left\vert \mu
\right\vert$. This is an interesting model because it is known that the
Klein-Gordon equation with the imaginary mass can be mapped to the telegraph
equation which is an important equation in classical physics but the
corresponding quantum equation is not known.

It is also somewhat related to the dynamics of symmetry breaking. As is
shown in Fig. \ref{fig1}, the state should move from the unstable local
maximum of an effective potential to the local minimum during the transition
between two phases: one is the restored phase (dashed line) and the other
the broken phase (solid line). It is considered that, at least, the early
stage of this time evolution will be approximately described by the
imaginary mass Klein-Gordon equation, for example, in the linear $\sigma$
model. One can easily see that there is no stationary solution in Eq.(\ref%
{epsilon}) since $\omega_\mathbf{k} $ becomes pure imaginary for $|\mathbf{q}%
_\mathbf{k}|<\left\vert \mu \right\vert $. That is, the Fock state vector
cannot be constructed as was discussed in Sec. \ref{chap:wf}.

There are several approaches to discuss the similar early-stage dynamics of
the phase transition \cite{boya}. As far as the authors know, most of
approaches are based on the semi-classical approximation of the full quantum
dynamics. Here we investigate this dynamics by solving the functional Schr%
\"{o}dinger equation.

\begin{figure}[t]
\includegraphics[scale=0.3]{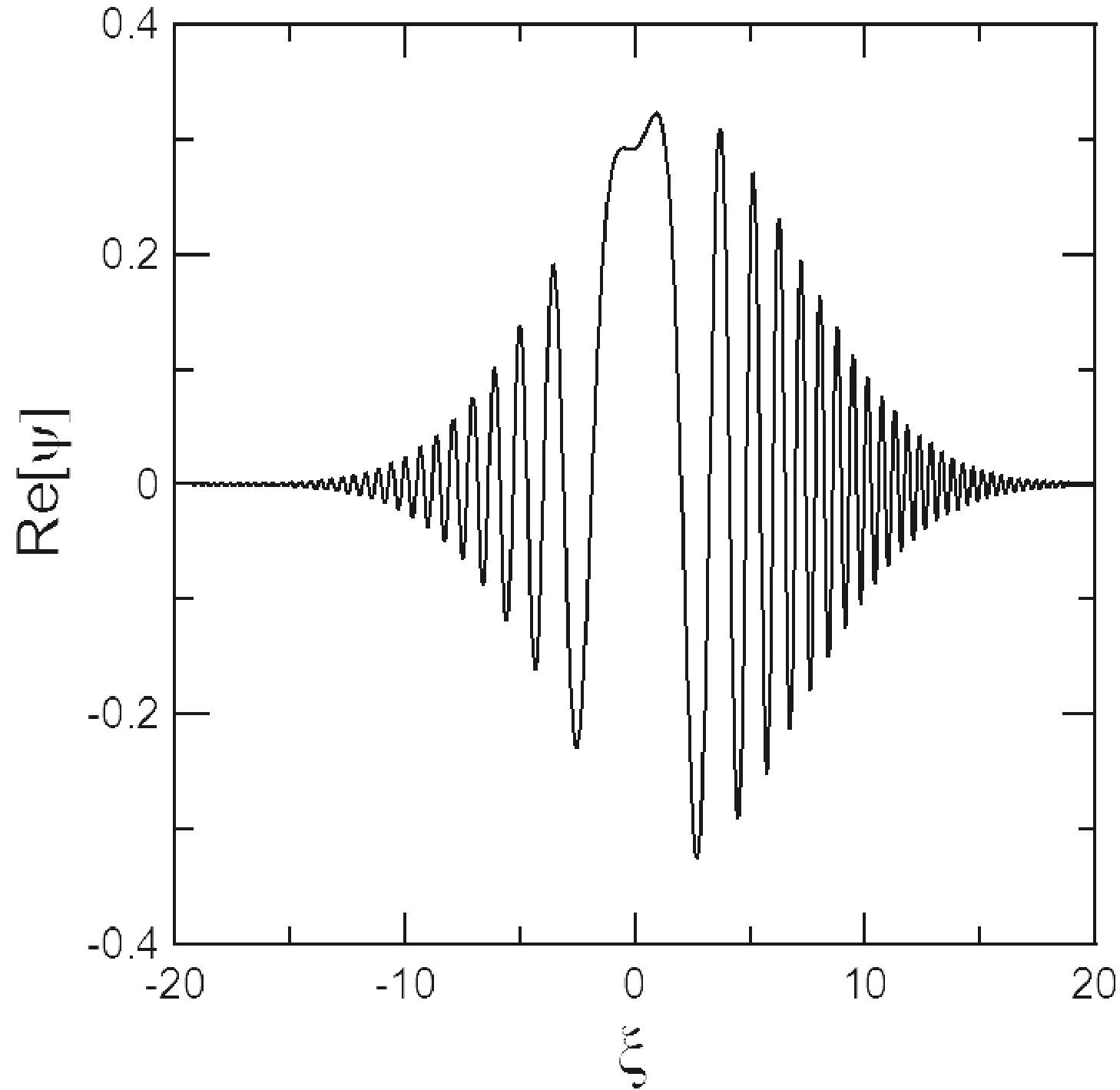} \includegraphics[scale=0.3]{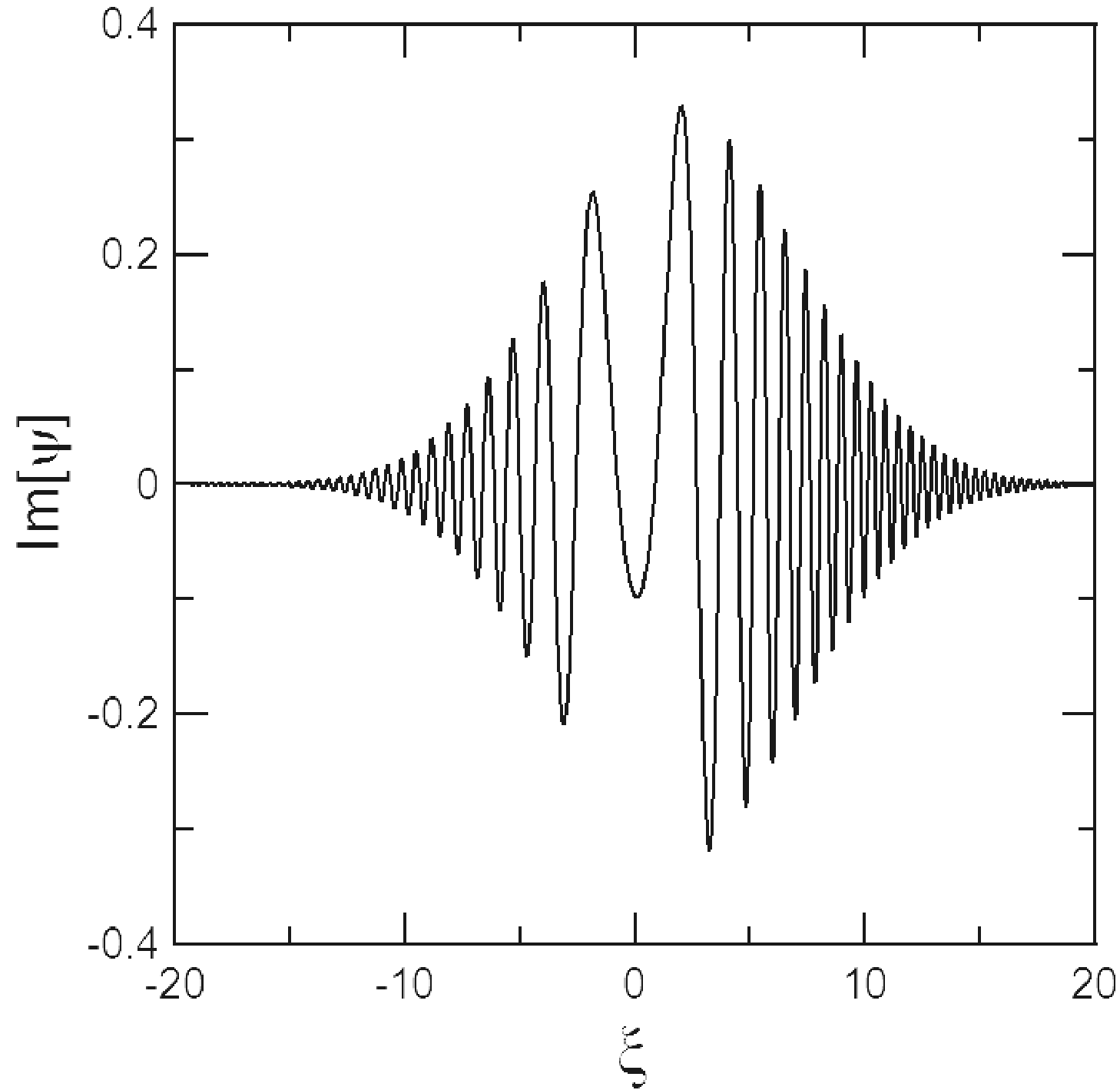}
\caption{The real part (the left panel) and the imaginary part (the right
panel) of $\protect\psi (\protect\xi,\protect\tau )$ at $\protect\tau =1$.
The initial condition is given by Eq. (\protect\ref{ini}). }
\label{Refig}
\end{figure}

For the sake of simplicity, we ignore the $C_{I,\mathbf{k}}$ variable and
consider the solution for $C_{R,\mathbf{k}}$. We further drop the suffix $%
\mathbf{k}$ since the adimensional form of the equation is universal for all 
$\mathbf{k}$ values. Then the functional Schr\"{o}dinger equation to be
solved is 
\begin{equation}
i\partial _{\tau }\psi (\xi )=\partial _{\xi }^{2}\psi (\xi )+\xi ^{2}\psi
(\xi ),
\end{equation}%
with $\tau =ct\sqrt{\left\vert \mu \right\vert ^{2}-\mathbf{q}_\mathbf{k}^{2}%
}/2=\tilde{\omega}t/2$ and $\xi = C_{R,\mathbf{k}} \sqrt{\tilde{\omega}
(\Delta^3 \mathbf{k} )/(\hbar c^2)} $. This is nothing but the adimansional
Schr\"{o}dinger equation with the inverted harmonic oscillator potential.

As the initial condition, we consider a Gaussian distribution around $\xi =
1/2$, 
\begin{equation}
\psi (\xi ,0)=e^{-(\xi - 1/2)^{2}/2}.  \label{ini}
\end{equation}

\begin{figure}[t]
\includegraphics[scale=0.3]{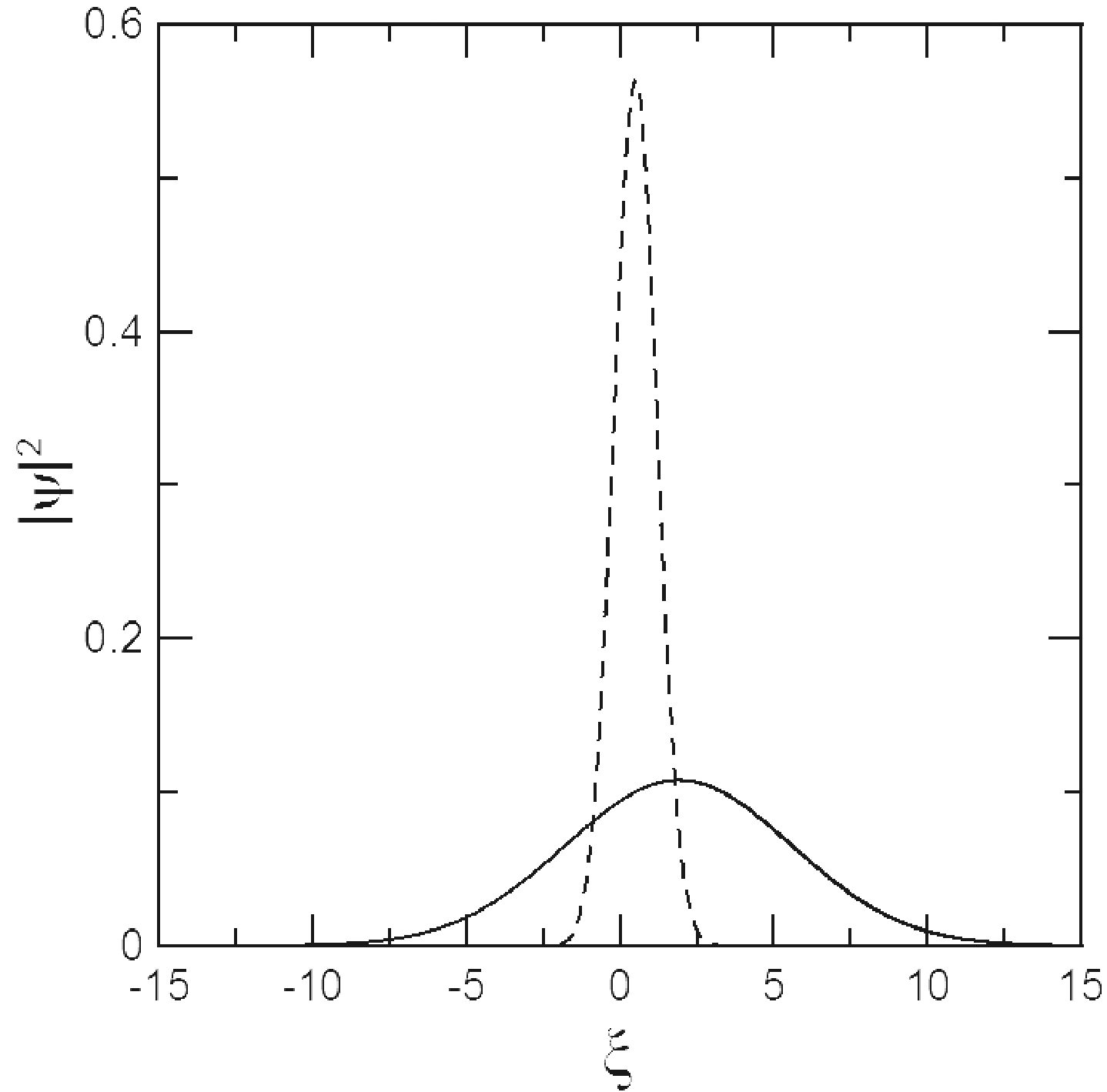}
\caption{The time evolution of the density $|\protect\psi(\protect\xi,%
\protect\tau)|^2$. The dashed and solid lines represent the initial
condition and the result at $\protect\tau=1$, respectively. }
\label{profig}
\end{figure}

The time evolutions of the real and imaginary parts of $\psi (\xi ,t)$ at $%
t=2/\tilde{\omega}$ are shown in the left and right panels of Fig. \ref%
{Refig}, respectively. One can see that the initial Gaussian distribution
diffuses exhibiting oscillation.

The corresponding configuration distribution $|\psi (\xi ,t)|^{2}$ is shown
in Fig. \ref{profig}. The initial density is denoted by the dashed line and
the result at $t=2/\tilde{\omega}$ is given by the solid line. The
configuration distribution shows a monotonic diffusion without oscillation,
and the initial distribution of the wave functional expands indefinitely.
The motion of the center of the wave functional is the same as the classical
motion of the field with the initial condition $\xi = 1/2$, and it seems
that the fluctuation of the field configuration evolves uniformly in
positions anterior and posterior of this classical path.

\subsection{Continuum limit}

\label{Sec:Cont}

We have developed the quantization in SVM by introducing the discretized
variables. The continuum limit of these variables are given by $%
L,N\rightarrow \infty $ keeping $\Delta^3 \mathbf{x} = (L/N)^3 \rightarrow 0$%
. Then the momentum sum appearing in our formulation is replaced by the
corresponding integral as 
\begin{equation}
\Delta ^{3}\mathbf{k}\sum_{\mathbf{k}} \longrightarrow \int d^{3}\mathbf{k}.
\end{equation}
And the commutation relations in this limit are expressed as 
\begin{eqnarray}
&&[a_{\mathbf{k}},a_{\mathbf{k^{\prime }}}^{\dagger }]=\delta _{\mathbf{k},%
\mathbf{k}^{\prime }}^{(3)}\longrightarrow \lbrack a(\mathbf{k}),a^{\dagger}(%
\mathbf{k^{\prime }})]=\delta ^{(3)}(\mathbf{k}-\mathbf{k}^{\prime }), \\
&&[b_{\mathbf{k}},b_{\mathbf{k^{\prime }}}^{\dagger }]=\delta _{\mathbf{k},%
\mathbf{k}^{\prime }}^{(3)}\longrightarrow \lbrack b(\mathbf{k}),b^{\dagger}(%
\mathbf{k^{\prime }})]=\delta ^{(3)}(\mathbf{k}-\mathbf{k}^{\prime }),
\end{eqnarray}%
where 
\begin{eqnarray}
\lim_{V\rightarrow \infty }\sqrt{\frac{1}{\Delta ^{3}\mathbf{k}}}a_{\mathbf{k%
}} &=&a(\mathbf{k}), \\
\lim_{V\rightarrow \infty }\sqrt{\frac{1}{\Delta ^{3}\mathbf{k}}}b_{\mathbf{k%
}} &=&b(\mathbf{k}),
\end{eqnarray}%
and 
\begin{equation}
\lim_{V\rightarrow \infty }\frac{1}{\Delta ^{3}\mathbf{k}}\delta _{\mathbf{k}%
,\mathbf{k}^{\prime }}^{\left( 3\right) }=\delta ^{(3)}(\mathbf{k}-\mathbf{k}%
^{\prime }).
\end{equation}%
One can confirm that the quantities calculated from the functional Schr\"{o}%
dinger equation such as the wave functional and the propagator have the
well-defined continuum limits by using the above results.

On the other hand, by introducing the functional derivative as 
\begin{equation}
\lim_{\Delta ^{3}\mathbf{k}\rightarrow 0}\frac{1}{\Delta ^{3}\mathbf{k}}%
\frac{\partial }{\partial C_{i,\mathbf{k}}} =\frac{\delta }{\delta C_{i,%
\mathbf{k}}},
\end{equation}%
the functional Schr\"{o}dinger equation is expressed as 
\begin{equation}
i\hbar \partial _{t}\psi (\underline{C}_{R},\underline{C}_{I},t)=%
\sum_{j=R,I}\int \frac{d^{3}\mathbf{k}}{(2\pi )^{3}}\left[ -\frac{\hbar
^{2}c^{2}}{2}\left( \frac{\delta }{\delta C_{j,\mathbf{k}}}\right) ^{2}+%
\frac{\omega^2_{\mathbf{k}}}{2c^2} C_{j,\mathbf{k}}^{2}\right] \psi (%
\underline{C}_{R},\underline{C}_{I},t) .
\end{equation}
This is the well-known form.
See also Refs. \cite{holland,sym,lu,kie,huang} for the functional
Schr\"{o}dinger equation.

Note that the higher order functional derivative such as $(\delta/\delta
C_{j,\mathbf{k}})^{2}$ appearing here is known to be able to provoke a
singular behavior. However, as is discussed in Refs. \cite{sym,lu}, this
singularity does not induce any problem for the calculations of various
observables when it is regularized by, for example, introducing the lattice.
Thus we need to keep in mind that the above continuum forms are defined in
the lattice representations per se. That is, the lattice representation
which was originally introduced to deal in stochastic variables is essential
to write down quantum dynamics in SVM.

There is another remark for the ultraviolet behavior of the single-particle
energy. In fact, because of $\mathbf{q}_{\mathbf{k}}$, the single-particle
energy $\omega_{\mathbf{k}} = c\sqrt{\mathbf{q}_{\mathbf{k}}^2 + \mu^2}$,
which agrees with the pole of the propagator (\ref{prop}), starts to deviate
from the usual expression $c\sqrt{\mathbf{k}^2 + \mu^2}$ near the momentum
cutoff 
\begin{equation}
|\mathbf{k}_{max}| = \frac{2\pi}{\Delta x}.
\end{equation}
Thus, to be consistent with the usual results in quantum field theory, 
$\Delta x$ must be sufficiently small, even when we calculate a
certain finite quantity.

\section{Concluding remarks}

\label{chap:last}

The formulation of the stochastic variational method was extended so as to apply it to 
the field quantization of the bosonic fields.

For this, we first introduced spatial lattice and define the field stochastic variables on 
these grid points.
We further showed that any unitary equivalent variable can be used for the SVM quantization.
In particular, the basis which diagonalizes the Laplacian operator is more convenient and 
it corresponds to the Fourier transform of a field when and only when the continuum limit is taken.

Dynamics of the quantized field in this scheme is determined by the Euler equation in the
functional space of the field configuration, and the
quantum effect is expressed by the functional quantum potential. 
This equation is further rewritten in the form of the functional Schr\"{o}dinger equation 
which can be obtained even in the canonical quantization \cite{holland,sym,lu,kie,huang}. 
The noise intensity is a parameter of the SVM quantization and determined by calculating
the single-particle energy defined by the Lagrange transform of the
stochastic Lagrangian. 
In this way, we can conclude that the time evolution of the wave
functional \textit{must} be induced by the Hamiltonian operator.

Differently from the particle systems, we need to introduce a Wiener process which has 
correlations between the different grid points 
(or discretized momentum), and the correlations are proportional to the Dirac delta function in the continuum limit.
This Wiener process leads to the second order functional derivatives in the functional Schr\"{o}dinger equation. 
It is known that such a term provokes a singular behavior but does not induce any problem in the calculations 
of observables as is discussed as is discussed in Ref. \cite{sym,lu}.
Thus it becomes clear that the SVM quantization with such a Wiener process still has a continuum limit.
Because of this singularity, 
we should interpret that the functional Schr\"{o}dinger equation is defined as the continuum limit of the 
the lattice representation.

The Fock state vectors are given by the stationary solutions of the
functional Euler equation or, equivalently, the eigenstates of the
functional Schr\"{o}dinger equation. By using the stochastic Noether
theorem, we further confirmed that the anti-particle degrees of freedom are
correctly reproduced on this Fock space. Thus we conclude that SVM
functions as the quantization scheme not only particles but also fields.

The purpose of the present paper is to formulate the SVM field
quantization so as to reproduce the results of the canonical quantization.
Nevertheless, we found that the SVM quantization has an advantage in the
definition of the quantized Noether charge because stochastic variables are
commutable and there is no ambiguity for the ordering of operators which is 
solved by introducing, for example, the normal ordering product.

It is known that, differently from quantum mechanics, infinitely many
unitary non-equivalent Fock spaces are contained in quantum field theory and
hence it is not trivial which one should be chosen. See, for example, Refs. 
\cite{unr1,unr2}. So far, this choice of the Fock space is known only in the
scattering theory and in the case where Landau's Fermi-liquid ansatz is
applicable. While it causes several difficulties, the presence of these
unitary non-equivalent Fock spaces manifests a rich structure of vacua in
quantum field theory, such as symmetry breaking. We confirmed that the
unitary non-equivalent representations appear in the continuum limit in our
formulation. See Appendix \ref{app:neur} for details.

To establish SVM as the alternative quantization scheme, the applicability
to the gauge field and the Dirac field should be studied. 
The application to thr gauge field is discussed in the Part II of this series of papers.
On the other hand, our formulation
may be considerably modified in the application to the
fermionic field. For example, the Wiener process used so far in the SDEs may
not be applicable in this case, because the stochastic properties of the
Dirac equation seems to be the Poisson process as is discussed in Refs. \cite%
{kac1,kac2,kac3}.

The stochastic quantization by Parisi and Wu is another quantization method
based on the stochastic dynamics \cite{parisi}. The principal difference
between SVM and Parisi-Wu comes from the origin of the noise. In the
stochastic quantization of Parisi-Wu, an additional time dimension is
introduced to quantize systems and the stochastic dynamics is considered for
this time. Besides this stochastic quantization, there are various
quantization schemes introducing an additional dimension \cite{mueller}. 
More detailed comparison with such approaches is an issue in the future.

\vspace{1.5cm}

Authors acknowledge the financial support of CNPq. PRONEX and FAPERJ.


\appendix

\section{Eigenvectors of Laplacian Matrix}

\label{app}

Let us consider the following eigenvalue problem, 
\begin{equation}
\Delta _{x}\underline{u}^{\left( n\right) }=\lambda _{n}\underline{u}%
^{\left( n\right) },
\end{equation}%
where $n=0,\cdots ,N-1$. As the eigenvector, we assume the following form, 
\begin{equation}
\underline{u}=\left( 
\begin{array}{c}
e^{i\alpha } \\ 
e^{i2\alpha } \\ 
\vdots \\ 
e^{iN\alpha }%
\end{array}%
\right)  \label{Ansatz}
\end{equation}%
with the periodic boundary condition as%
\begin{equation}
e^{i\left( N+l\right) \alpha }=e^{il\alpha },
\end{equation}%
for any integer $l$. Then $\alpha$ is determined by 
\begin{equation}
\alpha _{n}=\frac{2\pi }{N}n,
\end{equation}
where $n = 0, \cdots, N-1$. Substituting this assumption, the eigenvalue
equation gives the eigenvalue as 
\begin{equation}
\lambda _{n}=-\left[ \frac{2}{\Delta x}\sin \left( \frac{k_{n}\Delta x}{2}%
\right) \right] ^{2},
\end{equation}%
with 
\begin{equation}
k_{n}=\frac{2\pi }{L}n.
\end{equation}

Note that the pair of eigenvalues $\lambda _{n}$ and $\lambda _{N-n}$ are
degenerated, 
\begin{equation}
\lambda _{N-n}=\lambda _{n}.
\end{equation}%
Thus, for the sake of convenience, we assign a number to the eigenvectors as
follows, 
\begin{equation}
\ \underline{u}^{\left( n\right) }=\sqrt{\frac{1}{\pi }}\left( 
\begin{array}{c}
\sin \left( k_{n}x_{0}\right) \\ 
\sin \left( k_{n}x_{1}\right) \\ 
\vdots \\ 
\sin \left( k_{n}x_{N-1}\right)%
\end{array}%
\right) ,\ \underline{u}^{\left( n+1\right) }=\sqrt{\frac{1}{\pi }}\left( 
\begin{array}{c}
\cos \left( k_{n}x_{0}\right) \\ 
\cos \left( k_{n}x_{1}\right) \\ 
\vdots \\ 
\cos \left( k_{n}x_{N-1}\right)%
\end{array}%
\right) ,\   \label{real-u's}
\end{equation}%
where $n=1,3,..,N-2$ and the corresponding eigenvalues are degenerated,%
\begin{equation}
\lambda _{n}=\lambda _{n+1}.
\end{equation}%
For $n=0,$ we set 
\begin{equation}
\underline{u}^{\left( 0\right) }=\sqrt{\frac{1}{2\pi }}\left( 
\begin{array}{c}
1 \\ 
1 \\ 
\vdots \\ 
1%
\end{array}%
\right) .
\end{equation}
These eigenvectors form an orthogonal basis,%
\begin{equation}
\left( \underline{u}^{(n)}\ast \underline{u}^{(l)}\right) _{ x}=\frac{1}{%
\Delta k}\delta _{n,l},  \label{uiuj}
\end{equation}%
with the definition of the scalar product, 
\begin{equation}
\left( \underline{u}^{(n)}\ast \underline{u}^{(l)}\right) _{\Delta x}\equiv
\Delta x\sum_{m=0}^{N-1}\left( u^{(n)}\right) _{m}\left( u^{(l)}\right) _{m}.
\end{equation}

With this basis, any vector $\underline{f}$ in the $x$-space can be expanded
as 
\begin{equation}
\underline{f}=\Delta k\sum_{n=0}^{N-1}c_{n}\underline{u}^{(n)},  \label{y=Cu}
\end{equation}%
with%
\begin{equation}
c_{n}=\left( \underline{u}^{(n)}\ast \underline{f}\right) _{ x}.  \label{C}
\end{equation}%
Then this vector $\underline{c}^T =\left( c_{0},c_{1},..,c_{N-1}\right) $
forms the conjugate vector of $\underline{f}$ in the $k$-space.

Substituting this expression into Eq.(\ref{y=Cu}), we obtain the following
relation, 
\begin{equation}
\sum_{n=0}^{N-1}\underline{u}^{\left( n\right) }\underline{u}^{\left(
n\right) T}=\frac{1}{\Delta k\Delta x}I,  \label{Complete}
\end{equation}%
where $I$ is the $\left( N\times N\right) $ identity matrix. If we introduce 
$\left( N\times N\right) $ matrix $O$ as 
\begin{equation}
O=\left( 
\begin{array}{ccc}
\underline{u}^{\left( 0\right) } & \cdots & \underline{u}^{\left( N-1\right)
}%
\end{array}%
\right) ,
\end{equation}%
Equations (\ref{uiuj}) and (\ref{Complete}) are expressed more compactly as, 
\begin{equation}
O^{T}O=OO^{T}=\frac{1}{\Delta k\Delta x}I.  \label{Orthogonal}
\end{equation}

With this matrix, the transform between $\underline{f}$ and $\underline{c}$
is expressed as 
\begin{equation}
\underline{c}= (\Delta x) O^{T} \underline{f},  \label{c=of}
\end{equation}
or 
\begin{equation}
\underline{f}= (\Delta k) O \underline{c}.  \label{y=oc}
\end{equation}

By writing $b_{0}=c_{0}/\sqrt{2\pi },$ $a_{n}=c_{2n-1}/\sqrt{\pi },\
b_{n}=c_{2n}/\sqrt{\pi }$, Eq. (\ref{y=oc}) is expressed as 
\begin{equation}
f_{l}=b_{0}\frac{2\pi }{L}+\frac{2\pi }{L}\sum_{n=1}^{(N-1)/2}\left\{
a_{n}\sin k_{n}x_{l}+b_{n}\cos k_{n}x_{l}\right\} .  \label{Fourier}
\end{equation}%
Therefore, in the limit of $N\rightarrow \infty$, Eqs. (\ref{c=of}) and (\ref%
{y=oc}) coincide with the Fourier series of a continuous function $f(x)$,
with, at most, a finite number of discontinuities. However, in our case, $%
\left\{ f_l = f(x_{l})\right\} $ are defined only on discrete grid points so
that we cannot use the right-hand side of Eq.(\ref{Fourier}) to construct
continuous function for arbitrary $x$.

Another convenient basis can be used by introducing the (complex) linear
transformation among the two degenerate eigenstates as 
\begin{eqnarray}
\underline{u}^{\left( n\right) \prime } &=&\frac{1}{\sqrt{2}}\left( 
\underline{u}^{\left( n\right) }+i\underline{u}^{\left( n+1\right) }\right) ,
\\
\underline{u}^{\left( -n\right) \prime } &=&\frac{1}{\sqrt{2}}\left( 
\underline{u}^{\left( n\right) }-i\underline{u}^{\left( n+1\right) }\right)
\end{eqnarray}%
Then the new vector is expressed as 
\begin{equation}
\underline{u}^{\left( n\right) \prime }=\frac{1}{\sqrt{N}}\left( 
\begin{array}{c}
e^{-i\frac{N-1}{2}\alpha _{n}} \\ 
\vdots \\ 
1 \\ 
\vdots \\ 
e^{+i\frac{N-1}{2}\alpha _{n}}%
\end{array}%
\right) ,
\end{equation}%
for $n=-\frac{N-1}{2},\cdots ,0,\cdots ,\frac{N-1}{2}$.

By using this, Eq.(\ref{y=Cu}) is given by 
\begin{equation}
\underline{f}=\Delta k\sum_{n=-\left( N-1\right) /2}^{\left( N-1\right)
/2}c_{n}\underline{u}^{\left( n\right) \prime }.  \label{f=Cu-complex}
\end{equation}

These discretization scheme can be extended even in the higher dimensional
space in a similar manner by introducing the direct product space. For
example, for the two space-dimensional case $\left( x,y\right) $, the
eigenvector is expressed as 
\begin{equation}
\underline{u}^{\left( \mathbf{k}\right) }=\underline{u}_{x}^{\left( l\right)
}\otimes \underline{u}_{y}^{\left( m\right) }
\end{equation}%
where the index $\mathbf{k}$ denotes the combination of $(l,m)$. Then, the
normalization (\ref{uiuj}) is given by 
\begin{equation}
\left( \underline{u}^{\left( \mathbf{k}\right) }\ast \underline{u}^{\left( 
\mathbf{l}\right) }\right)_{\mathbf{x}} =\frac{1}{\left( \Delta k\right) ^{2}%
}\delta _{\mathbf{k,l}}.
\end{equation}

By using this basis, a function $f(\mathbf{x},t)$ in the two spatial
dimension is expanded as 
\begin{equation}
\underline{f}=\left( \Delta k\right) ^{2}\sum_{\mathbf{k}}c_{\mathbf{k}}%
\underline{u}^{\left( \mathbf{k}\right) }.
\end{equation}%
On the other hand, the spatial gradient matrices are extended as 
\begin{eqnarray}
\mathbf{\nabla }_{+} &=&\left( \nabla _{+}^{\left( x\right) }\otimes
I^{(y)}\right) \mathbf{e}_{x}+\left( I^{(x)}\otimes \nabla _{+}^{\left(
y\right) }\right) \mathbf{e}_{y}, \\
\mathbf{\nabla }_{-} &=&\left( \nabla _{-}^{\left( x\right) }\otimes
I^{(y)}\right) \mathbf{e}_{x}+\left( I^{(x)}\otimes \nabla _{-}^{\left(
y\right) }\right) \mathbf{e}_{y},
\end{eqnarray}%
where $\mathbf{e}_{x}$ and $\mathbf{e}_{y}$ are the spatial unit vector for
the $x$ and $y$ directions, respectively. Then the corresponding Laplacian
matrix is expressed by 
\begin{equation}
\mathbf{\Delta }_{\mathbf{x}}=\nabla _{+}^{\left( x\right) }\nabla
_{-}^{\left( x\right) }\otimes I^{(y)}+I^{(x)}\otimes \nabla _{+}^{\left(
y\right) }\nabla _{-}^{\left( y\right) }.
\end{equation}

\section{Quantization in $\mathbf{x}$-Space Representation}

\label{app:x}

We can apply the SVM quantization directly to the field $\phi (\mathbf{x},t)$
defined on the discretized lattice $\mathbf{x}$-space, in the same way
discussed in Sec.II.

To introduce the SDEs for the real quantities, we re-express the stochastic
action Eq. (\ref{lag_x_1}) in terms of the real scalar field as 
\begin{eqnarray}
L_{SVM} &=& \frac{1}{2} \sum_{j=I,R} E\left[ \frac{1}{c^{2}}\frac{( D%
\underline{\widehat{\phi}}_{j}(t) * D\underline{\widehat{\phi}}%
_{j}(t))_{\Delta \mathbf{x}} + (\tilde{D}\underline{\widehat{\phi}}_{j}(t) * 
\tilde{D}\underline{\widehat{\phi}}_{j} (t) )_{\Delta \mathbf{x}}}{2} \right.
\nonumber \\
&& + (\underline{\widehat{\phi}}_{j} (t) * \mathbf{\Delta}_{\mathbf{x}} 
\underline{\widehat{\phi}}_{j} (t) )_{\Delta \mathbf{x}} -\mu ^{2}( 
\underline{\widehat{\phi}}_{j} (t) * \underline{\widehat{\phi}}_{j}
(t))_{\Delta \mathbf{x}} \Biggr] ,
\end{eqnarray}%
where 
\begin{eqnarray}
\widehat{\phi}_{\mathbf{x}} (t) &=& \frac{\widehat{\phi}_{R,\mathbf{x}} (t)
+ i\widehat{\phi}_{I,\mathbf{x}} (t) }{\sqrt{2}} , \\
(\underline{f} * \underline{h} )_{\Delta \mathbf{x}} &=& \Delta^3 \mathbf{x}
\sum_{\mathbf{x}} f_{\mathbf{x}} h_{\mathbf{x}},
\end{eqnarray}
with $\Delta^3 \mathbf{x}$ being $(\Delta x)^3 = (L/N)^3$.

As was done for the classical string, the SDEs used in the $\mathbf{x}$%
-space are related to the ones in the $\mathbf{k}$-space through the
orthogonal matrix which diagonalizes the three dimensional Laplacian $%
\mathbf{\Delta}_{\mathbf{x}} $. Then, we obtain 
\begin{eqnarray}
d\widehat{\phi}_{i,\mathbf{x}}(t) &=& u_{i,\mathbf{x}}( \underline{\widehat{%
\phi}}_{R}(t),\underline{\widehat{\phi}}_{I}(t),t)dt+ \sqrt{\frac{2\nu }{%
(\Delta x)^3}}d{W}_{i,\mathbf{x}}(t), \\
d\widehat{\phi}_{i,\mathbf{x}}(t) &=& \tilde{u}_{i,\mathbf{x}}( \underline{%
\widehat{\phi}}_{R}(t),\underline{\widehat{\phi}}_{I}(t),t)dt + \sqrt{\frac{%
2\nu }{(\Delta x)^3}} d\tilde{W}_{i,\mathbf{x}}(t),
\end{eqnarray}
where 
\begin{eqnarray}
E\left[ dW_{i,\mathbf{x}}(t)\right] &=& E\left[ d\tilde{W}_{i,\mathbf{x}}(t)%
\right] = 0, \\
E\left[ dW_{j,\mathbf{x^{\prime }}}(t)dW_{i,\mathbf{x}}(t)\right] &=& E\left[
d\tilde{W}_{j,\mathbf{x^{\prime }}}(t)d\tilde{W}_{i,\mathbf{x}}(t)\right] =
\delta_{jk}\delta _{\mathbf{x},\mathbf{x^{\prime }}}^{(3)}|dt|,
\end{eqnarray}%
Here the index $i$ denotes $R$ or $I$. There is no correlation between $%
dW_{i,\mathbf{x}}(t)$ and $d\tilde{W}_{i,\mathbf{x}}(t)$.

The configuration distribution is then defined by 
\begin{equation}
\rho (\underline{\phi}_{R},\underline{\phi}_{I} ,t) = E\left[\prod_{\mathbf{x%
}} \delta ( \phi_{R,\mathbf{x}} - \widehat{\phi}_{R,\mathbf{x}} (t) ) \delta
( \phi_{I,\mathbf{x}} - \widehat{\phi}_{I,\mathbf{x}} (t) ) \right].
\end{equation}
The self-consistency condition is derived from the equivalence of the two
Fokker-Planck equations as 
\begin{equation}
u_{i,\mathbf{x}}(\underline{\phi}_{R},\underline{\phi}_{I} ,t) =\tilde{u}_{i,%
\mathbf{x}}(\underline{\phi}_{R},\underline{\phi}_{I},t) +\frac{2\nu }{%
\Delta^3 \mathbf{x}}\frac{\partial}{\partial {\phi_{i,\mathbf{x}}}}\ln \rho (%
\underline{\phi}_{R},\underline{\phi}_{I},t).
\end{equation}%
Using this condition, the Fokker-Planck equation is expressed as 
\begin{equation}
\partial_t \rho (\underline{\phi}_{R},\underline{\phi}_{I},t)= -
\sum_{j=I,R} \sum_{\mathbf{x}} \frac{\partial}{\partial \phi_{j,\mathbf{x}}}
\{ \rho (\underline{\phi}_{R},\underline{\phi}_{I},t) v_{j,\mathbf{x}}(%
\underline{\phi}_{R},\underline{\phi}_{I},t) \}.
\end{equation}
Here the mean functional velocity is defined by 
\begin{equation}
v_{i,\mathbf{x}}(\underline{\phi}_{R},\underline{\phi}_{I},t)=\frac{u_{i,%
\mathbf{x}}(\underline{\phi}_{R},\underline{\phi}_{I},t) +\tilde{u}_{i,%
\mathbf{x}}(\underline{\phi}_{R},\underline{\phi}_{I},t)}{2}.
\end{equation}

Then the functional Euler equation is obtained as the result of the
stochastic variation, 
\begin{eqnarray}
&& \left( \partial _{t}+\sum_{j=R,I} \sum_{\mathbf{x^{\prime }}} v_{j,%
\mathbf{x^{\prime }}}(\underline{\phi}_{R},\underline{\phi}_{I},t) \frac{%
\partial }{\partial \phi_{j,\mathbf{x^{\prime }}}} \right)v_{i,\mathbf{x}}(%
\underline{\phi}_{R},\underline{\phi}_{I},t)  \nonumber \\
&& - \frac{2\nu ^{2}}{(\Delta^3 \mathbf{x})^2} \frac{\partial }{\partial
\phi_{i,\mathbf{x}}} \left\{\rho ^{-1/2}(\underline{\phi}_{R},\underline{\phi%
}_{I},t) \sum_{j=R,I} \sum_{\mathbf{x^{\prime }}}\left( \frac{\partial}{%
\partial \phi_{j,\mathbf{x^{\prime }}}}\right)^{2}\rho ^{1/2} (\underline{%
\phi}_{R},\underline{\phi}_{I},t)\right\}  \nonumber \\
&=& c^2 ( (\mathbf{\Delta}_{\mathbf{x}}- \mu^2)\underline{\phi}_{i})_{%
\mathbf{x}}.  \label{eq-v-kg}
\end{eqnarray}

This is expressed in the form of the functional Schr\"{o}dinger equation, 
\begin{equation}
i\hbar \partial _{t}\Psi (\underline{\phi}_{R},\underline{\phi}_{I},t)= 
\frac{\Delta^3 \mathbf{x}}{2} \sum_{i=I,R} \sum_{\mathbf{x}} \left[ -\frac{%
\hbar ^{2}c^{2}}{(\Delta^3 \mathbf{x})^2} \frac{\partial^{2}}{\partial
\phi_{i,\mathbf{x}}^2} - \phi_{i,\mathbf{x}} \mathbf{\Delta}_{\mathbf{x}}
\phi_{i, \mathbf{x}} + \mu ^{2} \phi^2_{i,\mathbf{x}} \right] \Psi (%
\underline{\phi}_{R},\underline{\phi}_{I},t).
\end{equation}
Here the wave functional is given by 
\begin{equation}
\Psi (\underline{\phi}_{R},\underline{\phi}_{I},t)=\sqrt{\rho (\underline{%
\phi}_{R},\underline{\phi}_{I},t)}e^{i\theta (\underline{\phi}_{R},%
\underline{\phi}_{I},t)},
\end{equation}%
where 
\begin{equation}
v_{i,\mathbf{x}}(\underline{\phi}_{R},\underline{\phi}_{I},t)= \frac{2\nu}{%
\Delta^3 \mathbf{x}} \frac{\partial}{\partial \phi_{i,\mathbf{x}}} \theta (%
\underline{\phi}_{R},\underline{\phi}_{I},t).
\end{equation}

The vacuum wave functional is 
\begin{equation}
\Psi _{0}(\underline{\phi}_{R},\underline{\phi}_{I}) = N\exp \left\{ -\frac{1%
}{2\hbar c}\sum_{j=I,R} \int d^3 \mathbf{x} \phi_{j}(\mathbf{x}) \sqrt{-%
\mathbf{\Delta}_{\mathbf{x}} +\mu ^{2}} \phi_{j} (\mathbf{x}) \right\} ,
\end{equation}%
where 
\begin{equation}
N= \left( det \left[ \frac{\Delta^3 \mathbf{x} \sqrt{-\mathbf{\Delta}_{%
\mathbf{x}} +\mu ^{2}}}{\hbar c\pi } \right] \right) ^{1/2}.
\end{equation}
Here the operation of the operator $\sqrt{-\mathbf{\Delta}_{\mathbf{x}} +\mu
^{2}}$ is defined in the orthogonal basis of $\mathbf{\Delta}_{\mathbf{x}} $%
. One can easily check that this solution is the same as the functional
vacuum wave function obtained in the $\mathbf{k}$-space, (\ref{vac}), by
noting that $\phi_{i,\mathbf{x}}$ and $C_{i,\mathbf{k}}$ are connected
through the orthogonal transform.

\section{Unitary Non-equivalent Representations in Quantum Field Theory}

\label{app:neur}

The expression of the vacuum wave functional (\ref{vac}) resembles the
ground state wave function of the solution of the Schr\"{o}dinger equation
with the harmonic oscillator potential in quantum mechanics. In the quantum
mechanical case, the stationary states form the complete set and any time
evolution of state vectors can be expressed by the linear combination of
these states. However, this is not applicable in quantum field theory
because of the property of the unitary non-equivalent representation \cite%
{unr1,unr2}.

To see this, let us consider, for example, the following initial condition, 
\begin{equation}
\psi _{ini}(\underline{\phi}_{R},\underline{\phi}_{I}) =\Pi _{\mathbf{k}%
}\left( \frac{2\Omega _{\mathbf{k}}}{\pi \hbar c^{2}}\right)
^{1/2}e^{-\Omega _{\mathbf{k}}(C_{R,\mathbf{k}}^{2}+C_{I,\mathbf{k}%
}^{2})/(\hbar c^{2})},  \label{ini}
\end{equation}%
where $\Omega _{\mathbf{k}}=c\sqrt{\mathbf{q}_\mathbf{k}^{2}+M^{2}}$ with $%
M>m$. Suppose the time evolution of this initial state can be expressed by
the linear combination of the eigenstates which are obtained in Sec. \ref%
{chap:wf}. Then, for example, the overlap with the vacuum wave functional is
given by 
\begin{eqnarray}
\int D[C_{R}]D[C_{I}]\psi _{ini}^{\ast }(\underline{\phi}_{R},\underline{\phi%
}_{I})\psi _{0}(\underline{\phi}_{R},\underline{\phi}_{I}) &=&\Pi _{\mathbf{k%
}}\frac{2\sqrt{\Omega _{\mathbf{k}}\omega _{\mathbf{k}}^{m}}}{\omega _{%
\mathbf{k}}+\Omega _{\mathbf{k}}}  \nonumber \\
&=&e^{-\frac{1}{(\Delta k)^{3}}\int d^{3}\mathbf{k}\ln \frac{\omega _{%
\mathbf{k}}+\Omega _{\mathbf{k}}}{2\sqrt{\Omega _{\mathbf{k}}\omega _{%
\mathbf{k}}}}}=0.
\end{eqnarray}%
In the last line, we take the limit of $V\rightarrow \infty $ noting that $%
\left( \omega _{\mathbf{k}}+\Omega _{\mathbf{k}}\right) /2\sqrt{\Omega _{%
\mathbf{k}}\omega _{\mathbf{k}}}\geq 1$. Similarly, we can show that there
is no overlap between $\psi _{ini}(\underline{\phi}_{R},\underline{\phi}%
_{I}) $ and any Fock state vector constructed from $\psi _{0}(\underline{\phi%
}_{R},\underline{\phi}_{I})$. Thus, the time evolution should be calculated
by directly solving the functional Schr\"{o}dinger equation (\ref{sch_eq_am}%
) with the initial state (\ref{ini}).

As another example, let us consider a uniform translation of the amplitude
by a $c$-number, say $\phi _{0},$%
\begin{equation}
\phi \left( \mathbf{x}\right) \rightarrow \phi \left( \mathbf{x}\right)
+\phi _{0},
\end{equation}%
which is a coherent transform and often appears to introduce the spontaneous
breaking of symmetry. In the $\mathbf{k}$-space representation, this
corresponds to a shift by a amount of $\phi _{0}$ in the $\mathbf{k}=0$
mode. It is easy to verify that the overlap between the original and shifted
vacua is given by%
\begin{equation}
\exp \left\{ - \frac{1}{\Delta^3 \mathbf{k}} \frac{\mu }{2\hbar c}\phi
_{0}^{2}\right\}.
\end{equation}%
This disappears in the continuum limit, $\Delta^3 \mathbf{k} \rightarrow 0$.

\end{document}